\documentclass[runningheads]{llncs}

\PassOptionsToPackage{table}{xcolor}

\usepackage{eccv}

\usepackage{eccvabbrv}

\usepackage{graphicx}
\usepackage{booktabs}

\usepackage[accsupp]{axessibility}  

\usepackage[pagebackref,breaklinks,colorlinks,citecolor=eccvblue]{hyperref}

\usepackage{orcidlink}

\usepackage{amssymb}
\usepackage{array, multirow, bm}

\graphicspath{{Figure/}}
\newcommand{\tb}{\textbf}

\newcommand{\std}[1]{\tiny{$\pm{#1}$}}
\newcommand{\atick}{$\checkmark$}

\begin{document}

	\title{WiMANS: A Benchmark Dataset for\\WiFi-based Multi-user Activity Sensing} 
	
	
	\author{Shuokang Huang \and Kaihan Li \and Di You \and Yichong Chen \\ Arvin Lin \and Siying Liu \and Xiaohui Li \and Julie A. McCann}
	
	\authorrunning{S. Huang \textit{et al.}}
	
	\institute{Imperial College London, London SW7 2AZ, UK 			\\
	\email{\{s.huang21, j.mccann\}@imperial.ac.uk} 	\\
	\url{https://github.com/huangshk/WiMANS} }
	
	\maketitle
	
	%
	%
	%
	%
	%
	%
	%
	%
	%
	%
	%
	%
	%
	%
	%
	%
	%
	%
	%
	%
	\begin{abstract}
		WiFi-based human sensing has exhibited remarkable potential to analyze user behaviors in a non-intrusive and device-free manner, benefiting applications as diverse as smart homes and healthcare.
		However, most previous works focus on single-user sensing, which has limited practicability in scenarios involving multiple users.
		Although recent studies have begun to investigate WiFi-based multi-user sensing, there remains a lack of benchmark datasets to facilitate reproducible and comparable research.
		To bridge this gap, we present WiMANS, to our knowledge, the first dataset for multi-user sensing based on WiFi.
		WiMANS contains over 9.4 hours of dual-band WiFi Channel State Information (CSI), as well as synchronized videos, monitoring simultaneous activities of multiple users.
		We exploit WiMANS to benchmark the performance of state-of-the-art WiFi-based human sensing models and video-based models, posing new challenges and opportunities for future work.
		We believe WiMANS can push the boundaries of current studies and catalyze the research on WiFi-based multi-user sensing.
		\keywords{Human Sensing \and Multi-user \and WiFi \and Benchmark Dataset}
	\end{abstract}
	
	%
	%
	%
	%
	%
	%
	%
	%
	%
	%
	%
	%
	%
	%
	%
	%
	%
	%
	%
	%
	\section{Introduction}
		\label{section_introduction}

		\begin{table*}[t]
			\centering
			\tiny
			\caption{Comparison of public WiFi-based human sensing datasets. WiMANS is \textit{the first dataset} that involves multiple users performing different/identical activities simultaneously in each sample. ``Idt.'': Identity. ``Loc.'': Location. ``Act.'': Activity.}
			\begin{tabular}{l@{\;\;}c@{\;\;}c@{\;\;}c@{\;\;}c@{\;\;}c@{\;\;}c@{\;\;}c@{\;\;}c@{\;\;}c@{\;\;}c}
				\toprule
				\multirow{2}*{\tb{Dataset}}						& \tb{\# of Users}	& \tb{\# of}	& \tb{\# of}		& \tb{\# of}		& \tb{Sample}		& \tb{WiFi Band}	& \tb{Video}	& \multicolumn{3}{c}{\tb{Annotations}}	\vspace{-0.8mm}\\
				\cmidrule(){9-11}
				~												& \tb{per Sample} 	& \tb{Samples}	& \tb{Activities}	& \tb{Channels}	& \tb{Rate (Hz)}	& \tb{(GHz)}		& \tb{Data}		& Idt.	& Loc.	& Act.		\\
				\midrule
				Yousefi \textit{et al.} \cite{dataset_yousefi}	& 1					& 557			& 7				& 90				& 1000			& 5				& -					& -					& -			& \atick		\\
				SignFi \cite{dataset_signfi}					& 1					& 14280			& 276			& 90				& 12.5$\sim$200	& 5				& -					& -					& - 		& \atick	\\
				FallDeFi \cite{dataset_falldefi}				& 1					& 1070			& 19			& 90				& 1000			& 5				& -					& -					& -			& \atick	\\
				WiAR \cite{dataset_wiar}						& 1					& 4161			& 16			& 90				& 30			& 5				& -					& \atick			& - 		& \atick	\\
				ARIL \cite{dataset_aril}						& 1					& 1394			& 6				& 52				& -				& -				& -					& - 				& \atick	& \atick	\\
				Brinke \textit{et al.} \cite{dataset_brinke}	& 1					& 3749			& 6				& 270				& 20			& 2.4			& - 				& \atick			& - 		& \atick	\\
				RF-NET \cite{dataset_rf_net}					& 1					& 12000			& 6				& 60				& 100			& -				& -					& -					& -			& \atick	\\
				Baha \textit{et al.} \cite{dataset_baha}		& 1					& 9000			& 12			& 90				& 320			& 2.4			& -					& \atick			& -			& \atick	\\
				CSIDA \cite{dataset_csida}						& 1					& 2844			& 6				& 342				& 1000			& 5				& -					& \atick			& \atick  	& \atick	\\
				NTU-Fi \cite{dataset_ntufi}						& 1					& 2040			& 6				& 342				& 500			& 5				& -					& \atick			& -			& \atick	\\
				CPAR \cite{dataset_cpar}						& 1					& 560			& 7				& 64				& 1000			& 2.4			& -					& \atick			& -			& \atick	\\
				Widar \cite{dataset_widar}						& 1					& 54			& 2				& 90				& 2000			& 5				& -					& - 				& \atick	& \atick	\\
				Widar 2.0 \cite{dataset_widar2}					& 1					& 24			& 2				& 90				& 1000			& 5				& -					& - 				& \atick	& \atick	\\
				Widar 3.0 \cite{dataset_widar3}					& 1					& 271050		& 22			& 90				& 1000			& 5				& -					& \atick			& \atick	& \atick	\\
				Yang \textit{et al.} \cite{dataset_yang} 		& 1					& 1050			& 8				& 90				& 30			& 5				& -					& \atick			& \atick 	& \atick	\\
				Moshiri \textit{et al.} \cite{dataset_moshiri}	& 1					& 420			& 7				& 52				& 200			& 5				& -					& \atick			& -			& \atick	\\
				OPERAnet \cite{dataset_operanet}				& 0$\sim$1			& 6235			& 6				& 540				& 1600			& 5				& -					& \atick			& - 		& \atick	\\	
				MM-Fi \cite{dataset_mmfi}						& 1					& 1080			& 27			& 342				& 1000			& 5				& \atick			& \atick			& -			& \atick	\\
				SHARPax \cite{dataset_sharpax}					& 0$\sim$1			& 108			& 3				& 242$\sim$996		& 133			& 5				& -					& -					& -			& \atick	\\
				\midrule
				WiMANS (Ours)									& 0$\sim$5			& 11286			& 9				& 270				& 1000			& 2.4/5		& \atick			& \atick			& \atick 	& \atick	\\
				\bottomrule
			\end{tabular}
			\label{table_compare_dataset}
		\end{table*}
		
		%
		Recent years have witnessed the rapid progress of WiFi-based human sensing \cite{survey_2022_0,survey_2022_1,survey_2023}, which collects Channel State Information (CSI) from off-the-shelf WiFi devices to recognize human identities \cite{related_efficientfi,related_caution,related_wirelessid}, locations \cite{method_multiuser_1,dataset_widar2,related_three}, activities \cite{model_that,related_wigrunt,model_ablstm}, \textit{etc}.
		%
		It plays an increasingly important role in differing applications such as security monitoring \cite{app_security_0,app_security_1,app_security_2}, smart homes \cite{app_smarthome_0,app_smarthome_1}, and healthcare \cite{app_healthcare_0,app_healthcare_1,dataset_falldefi}.
		%
		Compared with cameras and on-body sensors, the use of WiFi CSI negates the necessity of filming users or attaching sensors to them \cite{survey_2021}.
		Such non-intrusive approaches can make human sensing widely available \cite{related_target}, satisfying the need to monitor users who do not want to be filmed or wear sensors.
		%
		WiFi-based human sensing is also robust to low-light or non-line-of-sight conditions, while using cameras for video-based analysis may be sensitive to low-light conditions and obstacles \cite{related_push}.
		%
		More importantly, CSI can be gathered from ubiquitous existing WiFi devices, enabling device-free human sensing \cite{survey_device_free} without the prerequisite of dedicated devices and particular deployments.
		
		%
		The principle of WiFi-based human sensing is that human activities essentially interfere with WiFi signals and lead to signal variations \cite{wifi_csi_0}.
		%
		Such variations are recorded in CSI, which thereby contains implicit features for human sensing. 
		%
		Since these human features typically intertwine with excessive noise \cite{wifi_csi_1}, it is impossible to interpret WiFi CSI patterns for human sensing straightforwardly.
		Therefore, various models have been proposed to learn features from the CSI for different human sensing tasks.
		%
		(1) For human identification, extensive literature has discussed the use of Multilayer Perceptrons (MLPs) \cite{dataset_ntufi}, Long Short-Term Memory (LSTM) \cite{related_wihi}, Convolutional
		Neural Networks (CNNs) \cite{related_caution,related_efficientfi,related_autofi}, CNN-LSTM hybrids \cite{related_push,related_wirelessid,model_clstm}, \textit{etc}.
		%
		(2) For human localization, prevalent models have been adopted, including Naive Bayes \cite{related_nb}, Auto-encoders \cite{related_radio_image}, LSTM \cite{related_three}, and CNNs \cite{model_cnn_1d}.
		%
		(3) WiFi-based human activity recognition (HAR) has drawn the greatest attention from researchers, applying models such as MLPs \cite{related_crosssense}, LSTM \cite{dataset_yousefi}, CNNs \cite{model_cnn_2d,related_pcnn,related_cnn_3d}, CNN-LSTM hybrids \cite{related_deepsense,related_lstm_cnn,related_gru}, Generative Adversarial Networks (GANs) \cite{related_gan}, attention-based bidirectional LSTM (ABLSTM) \cite{model_ablstm}, and Transformers \cite{related_transformer,model_that}.
		%
		
		%
		\begin{figure*}[t]
			\centering
			\begin{subfigure}{\linewidth}
				\centering
				\includegraphics[width=1\linewidth]{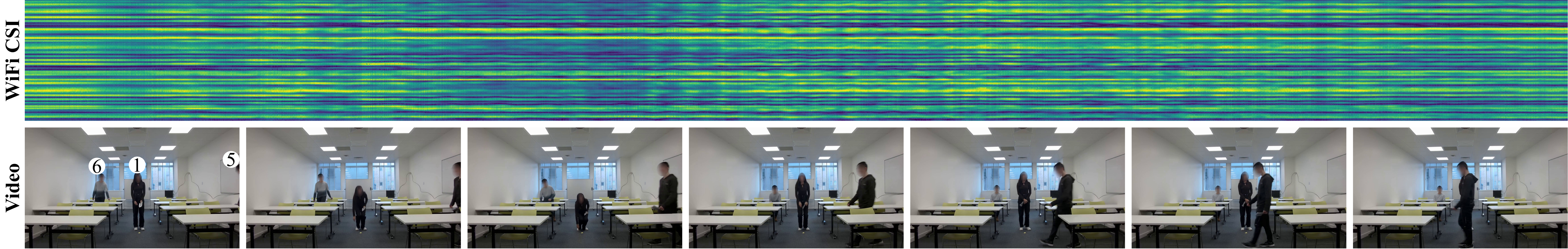}
				\caption{User 1: Picking Up. User 5: Walking. User 6: Sitting Down.}
				\label{figure_act_30_25}
			\end{subfigure}
			
			\begin{subfigure}{\linewidth}
				\centering
				\includegraphics[width=1\linewidth]{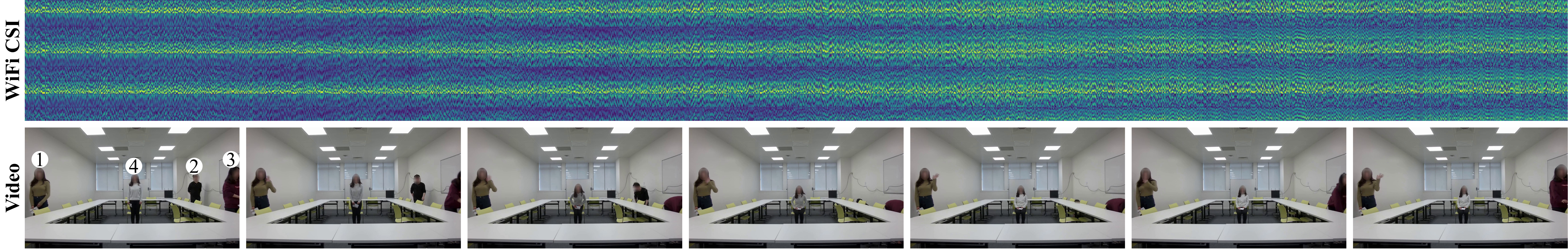}
				\caption{User 1: Waving. User 2: Lying Down. User 3: Picking Up. User 4: Sitting Down.}
				\label{figure_act_49_41}
			\end{subfigure}
			
			\begin{subfigure}{\linewidth}
				\centering
				\includegraphics[width=1\linewidth]{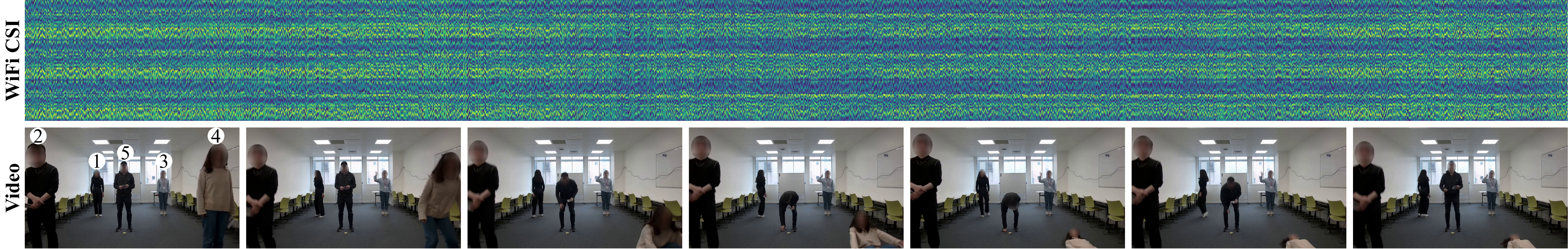}
				\caption{User 1: Rotation. User 2: Jumping. User 3: Waving. User 4: Lying Down. User 5: Picking Up.}
				\label{figure_act_88_30}
			\end{subfigure}
			\caption{Examples of WiFi CSI and videos in WiMANS, monitoring simultaneous activities performed by multiple users in various environments. (a) consists of 3 users in a classroom. (b) contains 4 users in a meeting room. (c) has 5 users in an empty room.}
			\label{figure_act}
		\end{figure*}
		
		%
		However, several critical drawbacks seriously hamper the practicability of WiFi-based human sensing.
		%
		(1) \textit{Single-user Limitation}: Existing public datasets only include a single user in each CSI sample, as shown in Table \ref{table_compare_dataset}.
		Consequently, most WiFi-based models only recognize the identity/location/activity of a single user in each recognition, but many practical scenarios indeed consist of multiple users simultaneously. 
		%
		Although recent studies \cite{method_multiuser_0,method_multiuser_1,method_multiuser_2,method_multiuser_3} have attempted to sense multiple users simultaneously, there remains \textbf{a lack of public datasets that enable WiFi-based multi-user sensing}.
		%
		(2) \textit{Insufficient Modalities and Annotations}: Most datasets collect CSI of a single WiFi band (\textit{i.e.}, 2.4 or 5 GHz) and do not incorporate synchronized videos.
		Such insufficiency disables the further study of unexplored tasks (\textit{e.g.}, pose estimation).
		Moreover, many datasets annotate samples for specific tasks (\textit{e.g.}, activity recognition), restricting the use of WiFi CSI in diverse sensing tasks.
		(3) \textit{Lack of Comprehensive Benchmarks}: Previous works mainly focus on novel models but few of them have provided benchmarks for these models.
		Specifically, no benchmark is available for multi-user activity sensing based on WiFi CSI.
		
		%
		To help resolve the above drawbacks, we present WiMANS, the first dataset that enables multi-user sensing based on WiFi CSI.
		%
		WiMANS collects 11286 CSI samples of dual WiFi bands (2.4 / 5 GHz), along with synchronized videos for reference, as shown in Fig. \ref{figure_act}.
		Each 3-second sample includes 0 to 5 users performing identical/different activities simultaneously, annotated with (anonymized) user identities, locations, and activities.
		Table \ref{table_compare_dataset} compares WiMANS with related datasets to highlight our novelty.
		%
		%
		Extensive experiments have been conducted to benchmark state-of-the-art WiFi-based models and video-based models.
		The main contributions and unique aspects of WiMANS are as follows:
		%
		\begin{enumerate}
			\item[\textbullet]{%
				We construct a WiFi-based multi-user activity sensing dataset, where each sample monitors simultaneous activities of multiple users.
				To the best of our knowledge, WiMANS is the first dataset that collects dual-band WiFi CSI and videos for multiple users in each sample. 
			}
			\item[\textbullet]{%
				%
				%
				WiMANS provides fine-grained annotations of user identities, locations, and activities to support various sensing tasks.
				The videos in WiMANS can further act as a reference for unexplored tasks (\textit{e.g.}, multi-user pose estimation).
				%
			}
			\item[\textbullet]{%
				%
				Benchmark experiments have been conducted to analyze the multi-user sensing performance of WiFi-based models and video-based models.
				This work provides the first benchmarks for WiFi-based multi-user identification, localization, and activity recognition.
				%
			}
		\end{enumerate}
		
	%
	%
	%
	%
	%
	%
	%
	%
	%
	%
	%
	%
	%
	%
	%
	%
	%
	%
	%
	%
	\section{Related Work}
		\label{section_related_work}
		\subsection{WiFi-based Human Sensing}
			%
			Human sensing with WiFi CSI is an increasingly promising alternative to traditional sensing technologies thanks to its non-intrusiveness, environmental robustness, and device-free merits.
			In WiFi-based human sensing, much effort has been devoted to three underlying yet distinct tasks: (1) human identification, (2) human localization, and (3) human activity recognition (HAR).
			
			%
			\textit{WiFi-based human identification} serves to determine user identities by learning biometric patterns from CSI.
			%
			To map CSI to identities, MLPs \cite{dataset_ntufi} simply feed all CSI values into fully connected layers, where excessive parameters lead to high complexity and poor generalization.
			%
			In contrast, LSTM \cite{related_wihi} regards CSI as sequences to learn temporal features.
			Despite the advanced performance than MLPs, LSTM is inefficient for lengthy sequences due to its step-by-step inputs.
			%
			To boost both effectiveness and efficiency, CNNs \cite{related_caution,related_efficientfi,related_autofi} have been utilized by dividing the CSI into diverse receptive fields to learn spatial features with convolution filters.
			%
			To combine the advantages of CNN and LSTM, recent works \cite{related_push,related_wirelessid,model_clstm} have proposed various CNN-LSTM hybrids, attaining the best performance for WiFi-based human identification.
			%
			%
			
			%
			\textit{WiFi-based human localization} estimates user locations to support human-computer interaction systems \cite{survey_2022_0}.
			%
			Initially, Naive Bayes \cite{related_nb}, a basic statistic model, has been leveraged for localization, showing acceptable performance.
			%
			Thereafter, Sparse Auto-encoder (SAE) \cite{related_radio_image} has been used to localize users based on CSI images, but its fully connected layers still lead to excessive parameters.
			%
			%
			LSTM \cite{related_three} and CNNs \cite{model_cnn_1d} have been further explored for localization, where CNN-1D \cite{model_cnn_1d} shows its superior capability for WiFi-based human localization. 
			
			%
			\textit{WiFi-based HAR} is gaining popularity for user behavior analysis, which is of finer granularity than identification and localization.
			Initial methods \cite{dataset_yousefi} based on handcrafted feature extraction have demonstrated inadequate ability to extract implicit features from CSI. 
			%
			Thereby, LSTM \cite{dataset_yousefi} and CNNs \cite{model_cnn_2d,related_cnn_3d,related_pcnn} have been extensively adopted and offer valuable insights on learning temporal and spatial features.
			%
			%
			CNN-LSTM hybrids \cite{related_deepsense,related_lstm_cnn,related_gru} further leverage the advantages of CNNs and LSTM to become predominant models for HAR.
			%
			%
			To tackle dynamic environments, GANs \cite{related_gan} have been used to augment CNNs with adversarial learning. 
			%
			Attention-based models \cite{ml_attention} also contribute to HAR \cite{related_transformer}, such as ABLSTM \cite{model_ablstm} which equips bidirectional LSTM with attention layers to enhance HAR performance.
			%
			%
			Recently, a two-stream convolution augmented transformer (THAT) \cite{model_that} has further integrated attention layers with multi-scale CNNs, achieving state-of-the-art performance for WiFi-based HAR.
			
			%
			However, all the above methods carry out single-user sensing only, while multi-user sensing based on WiFi CSI is more practical yet challenging.
			%
			Multiple users in an environment may occlude each other, challenging the resolution of models.
			The mutual interference between users also requires models to disentangle features of different users for effective sensing.
			Recent works \cite{method_multiuser_0,method_multiuser_1,method_multiuser_2,method_multiuser_3,method_multiuser_4,method_multiuser_5} have attempted to tackle these challenges and sense multiple users simultaneously, but they have not published their datasets and have not provided any benchmark. 
			%
			Such an obvious need motivates WiMANS as the first benchmark dataset that involves multiple users in each CSI sample.
			%
		
		%
		\subsection{Datasets for WiFi-based Human Sensing}
			High-quality annotated CSI datasets are essential for advancing WiFi-based human sensing research.
			Table \ref{table_compare_dataset} summarizes existing CSI datasets in comparison to WiMANS.
			At the beginning, Yousefi \textit{et al.} \cite{dataset_yousefi} collected a dataset in an office for WiFi-based HAR, after which various CSI datasets were released for HAR in different environments \cite{dataset_brinke,dataset_wiar,dataset_yang,dataset_moshiri}.
			Baha \textit{et al.} \cite{dataset_baha} further constructed a dataset under line-of-sight and non-line-of-sight conditions.
			Some datasets were presented for specific tasks, such as SignFi \cite{dataset_signfi} for sign language recognition, FallDeFi \cite{dataset_falldefi} for fall detection, and ARIL \cite{dataset_aril} for activity recognition and localization.
			For human tracking, Widar \cite{dataset_widar} and Widar 2.0 \cite{dataset_widar2} were gathered and annotated with user walking traces.
			%
			Some datasets contain changing scenarios, such as CPAR \cite{dataset_cpar} for different users and RF-NET \cite{dataset_rf_net} for varying environments.
			Similarly, CSIDA \cite{dataset_csida} and Widar 3.0 \cite{dataset_widar3} incorporate CSI of various domains (\textit{i.e.}, locations, orientations, environments, and users).
			OPERAnet \cite{dataset_operanet}, a multi-modal dataset, includes WiFi CSI as well as data from Passive WiFi Radar (PWR), Ultra-Wideband (UWB), and Kinect skeleton sensors.
			MM-Fi \cite{dataset_mmfi} is also a multi-modal dataset, containing WiFi CSI, video frames, depth frames, LiDAR point cloud, and mmWave radar point cloud.
			For benchmarking, Yang \textit{et al.} \cite{dataset_ntufi} collected the NTU-Fi dataset to compare WiFi-based human sensing models in terms of their performance, sizes, complexity, \textit{etc}.
			To analyze the impact of wireless parameters (\textit{e.g.,} bandwidth), Meneghello \textit{et al.} \cite{dataset_sharpax} constructed the SHARPax dataset to explore WiFi-based HAR under varying settings.
			
			%
			Despite these datasets, none of them enables simultaneous multi-user sensing, which is crucial for practical WiFi-based human sensing advancement, since real-life scenarios typically involve multiple users simultaneously. 
			To bridge this gap, we propose WiMANS, the first dataset for WiFi-based multi-user activity sensing, hoping to facilitate the next generation of human sensing based on WiFi.
			
	
	%
	%
	%
	%
	%
	%
	%
	%
	%
	%
	%
	%
	%
	%
	%
	%
	%
	%
	%
	%
	\section{WiMANS}
		\label{section_wimans}
		%
		WiMANS aims to gather WiFi CSI samples which monitor simultaneous activities of multiple users.
		Each CSI sample contains 0 to 5 users performing identical/different activities at the same time, as shown in Fig. \ref{figure_act}.
		%
		%
		Fine-grained annotations are provided for all samples, including user identities, locations, and activities.
		%
		Ultimately, WiMANS collects 11286 samples (over 9.4 hours) of dual-band WiFi CSI and synchronized videos. %
		
		%
		\subsection{Dataset Construction}
			\label{subsection_dataset_construction}
			%
			%
			\noindent\textbf{Hardware Setup.}
			WiFi CSI monitors the variations of wireless signals by which a transmitter propagates packets to a receiver.
			To collect WiFi CSI, we deploy two off-the-shelf computers (HP EliteDesk 800 G2 TWR) to serve as the transmitter and receiver.
			Each device is equipped with an Intel 5300 Network Interface Card and has the Linux 802.11n CSI tool \cite{experiment_80211} installed, following the previous works \cite{dataset_yousefi,dataset_widar3,dataset_operanet}.
			We set these devices to work in the monitor mode, enabling us to control the start and the end of each CSI sample.
			To exert the advantages of different WiFi bands, we collect dual-band WiFi CSI by setting devices to work on channel 12 for the 2.4 GHz band, and on channel 64 for the 5 GHz band.
			%
			Meanwhile, we use a monitor camera to capture synchronized videos.

			\noindent\textbf{Data Collection.}
				We employ the transmitter and receiver to collect a CSI sample in 3 steps:
				(1) The receiver listens to a WiFi channel and logs the CSI from all packets it receives;
				(2) The transmitter sends packets to the WiFi channel, and the users simultaneously perform designated activities;
				(3) The receiver stops logging and listening.
				%
				Following the previous works \cite{dataset_widar3,dataset_operanet}, we instruct users to perform activities in 3 seconds and thereby control the transmitter to send 3000 packets at a rate of 1000 packets per second.
				Regardless of packet loss, each 3-second CSI sample should have 3000 time steps under the sample rate of 1000 Hz.
				We will further analyze the issue of packet loss in Section \ref{subsection_dataset_statistic}.
				The transmitter and receiver individually have 3 antennas, and each pair of antennas uses 30 subcarriers for wireless communication.
				Thus, the dimension of CSI at each time step is 3$\times$3$\times$30, and the dimension of each CSI sample is 3000$\times$3$\times$3$\times$30.
				%
				%
				Along with each CSI sample, we also capture synchronized videos with a monitor camera for reference and unexplored tasks (\textit{e.g.}, multi-user pose estimation).
				Each 3-second video has 90 frames under the frame rate of 30 Hz, with 3 RGB channels and the frame resolution of 1920$\times$1080.
				Hence, the dimension of each video sample is 90$\times$3$\times$1920$\times$1080.

			\noindent\textbf{Data Synchronization.}
				During data collection, WiFi CSI is collected sample by sample, while long videos are recorded at the same time.
				Afterward, we synchronize WiFi CSI and videos by segmenting long videos into 3-second samples based on timestamps.
				Because the frame rate of videos is 30 Hz while the sample rate of CSI is 1000 Hz, we can synchronize WiFi CSI and video samples in 16.67 ms, which is unavoidable and tolerable for 3-second samples.
			
\begin{figure}[t]
				\centering
				\begin{subfigure}{0.95\linewidth}
					\centering
					\includegraphics[width=\linewidth]{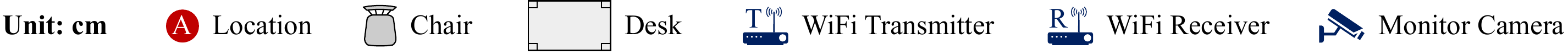}
				\end{subfigure}
				\vspace{1mm}
				
				\begin{subfigure}{0.28\linewidth}
					\centering
					\includegraphics[width=\linewidth]{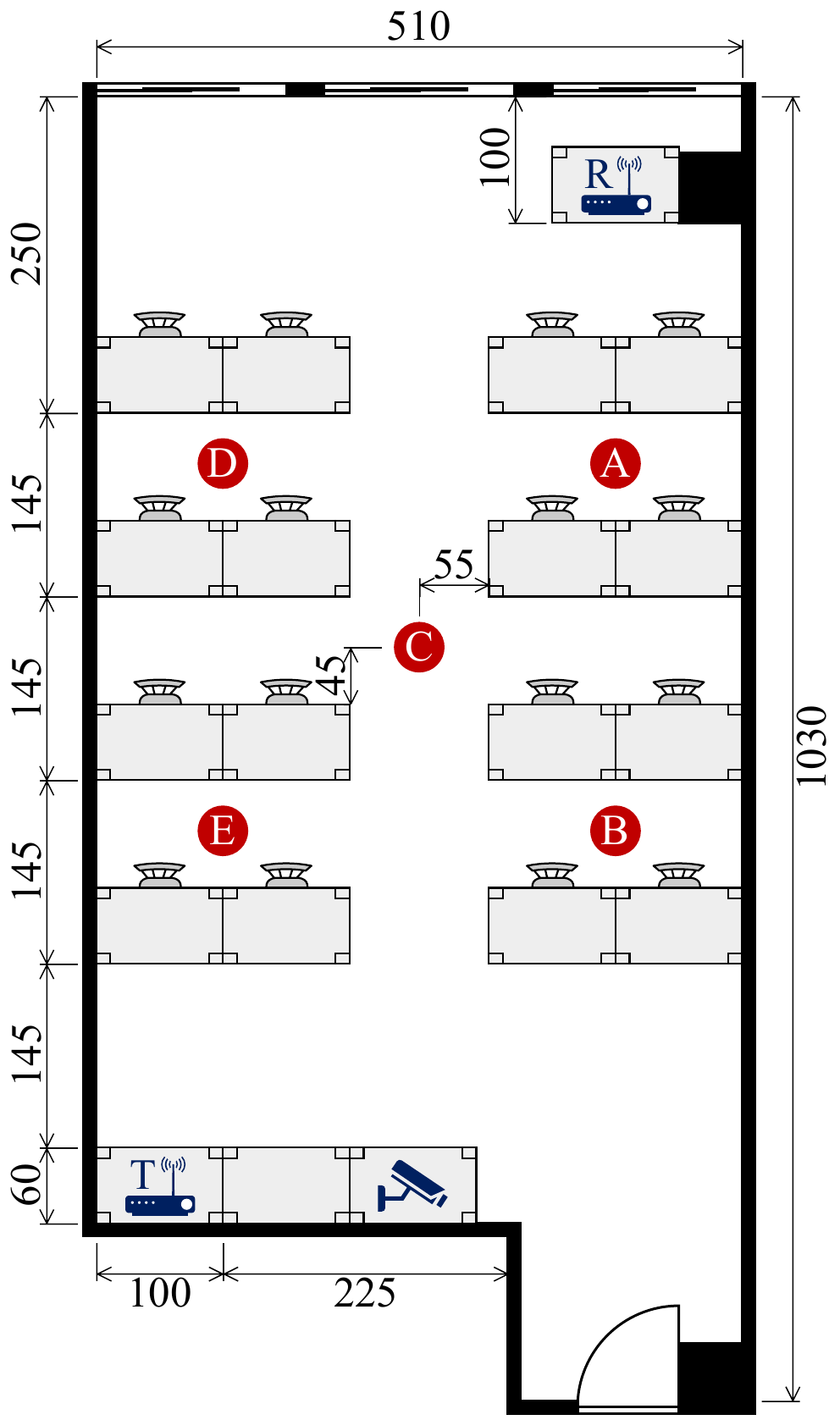}
					\caption{Classroom.}
				\end{subfigure}
				\hfill
				\begin{subfigure}{0.28\linewidth}
					\centering
					\includegraphics[width=\linewidth]{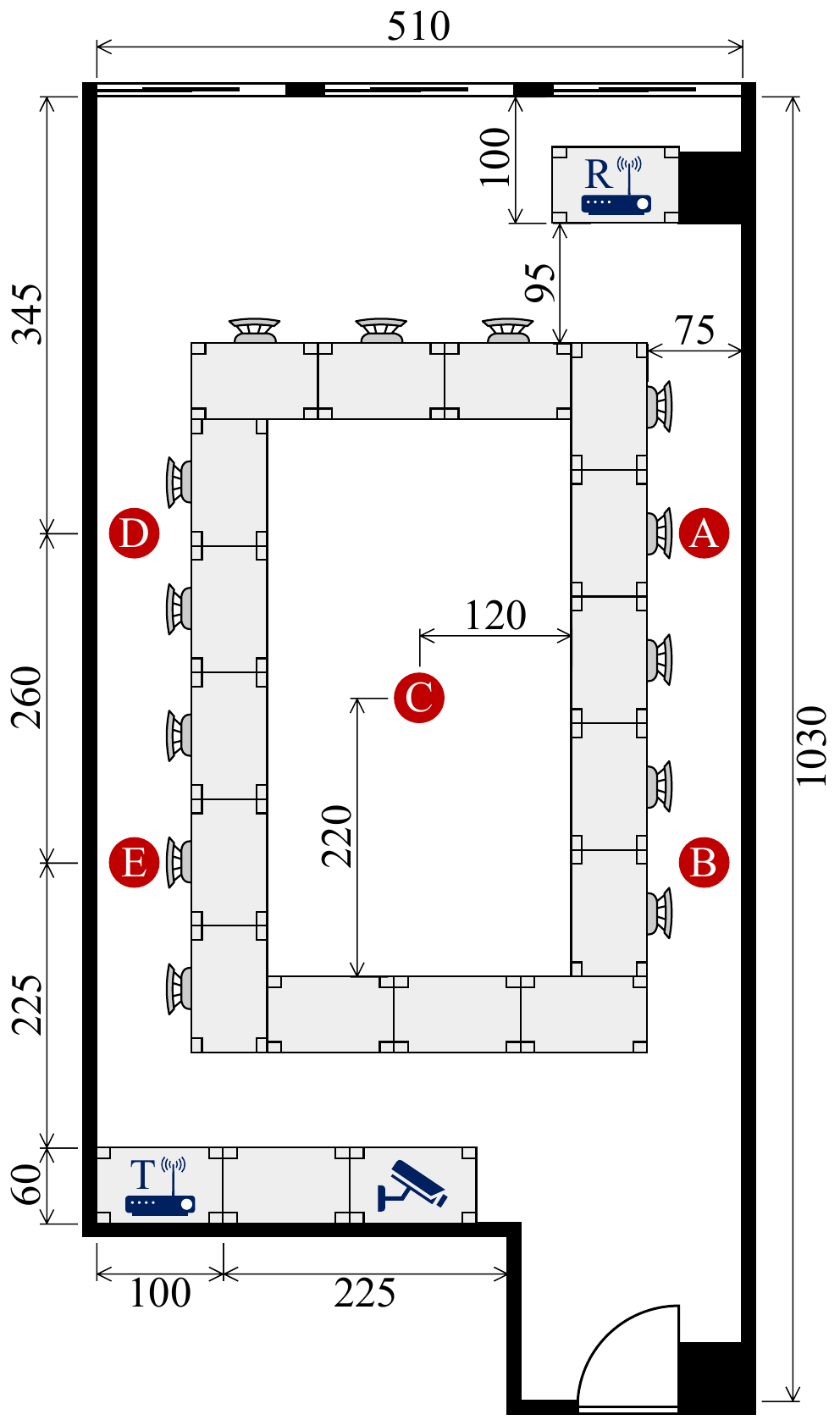}
					\caption{Meeting Room.}
				\end{subfigure}
				\hfill
				\begin{subfigure}{0.28\linewidth}
					\centering
					\includegraphics[width=\linewidth]{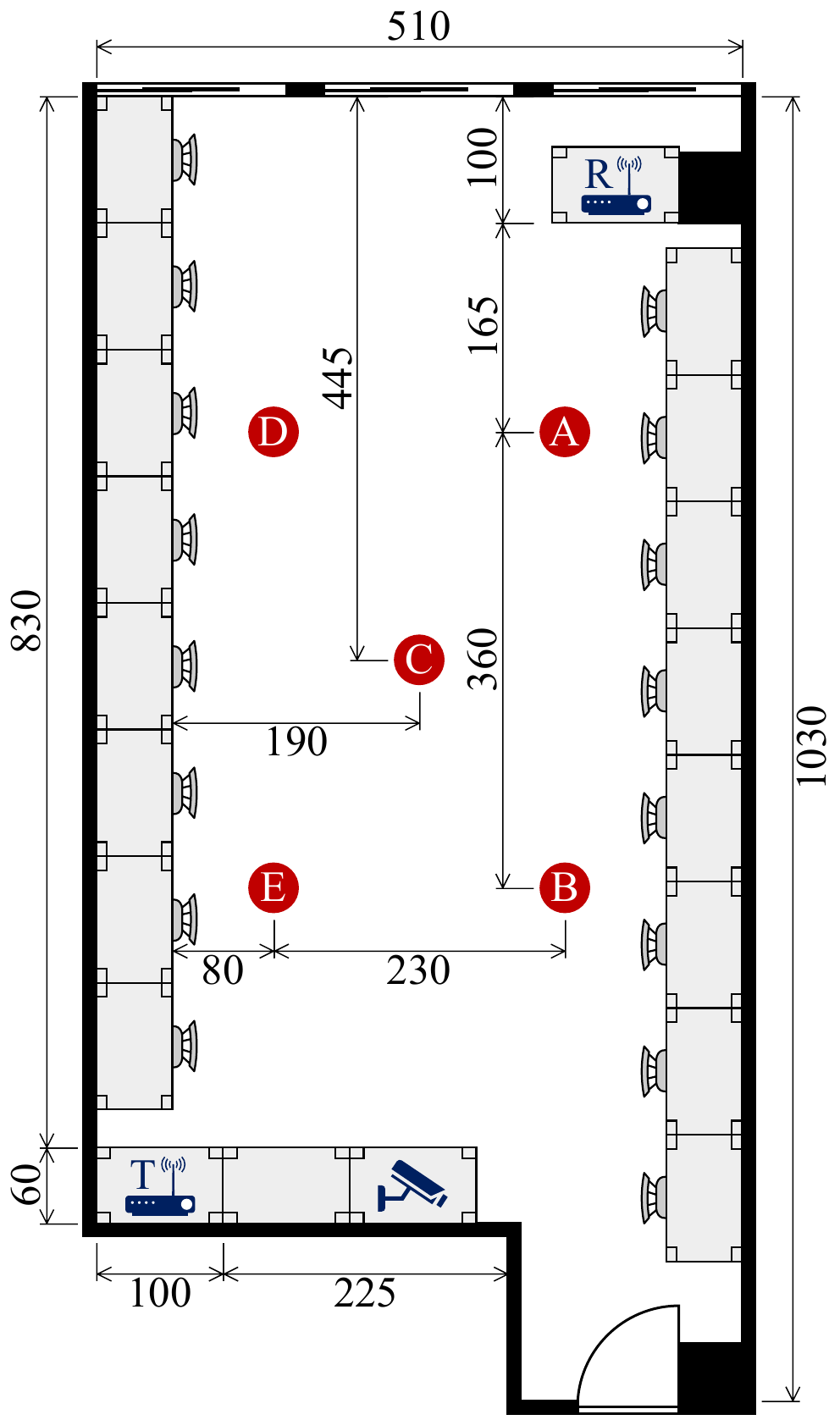}
					\caption{Empty Room.}
				\end{subfigure}
				\caption{Layouts of environments in WiMANS, where a transmitter and a receiver collect WiFi CSI, and a monitor camera captures synchronized videos for reference.}
				\label{figure_layout}
			\end{figure}

			%
			\noindent\textbf{Data Attributes.}
				WiMANS includes 9 activities which are representative in daily life, as discussed in the previous works \cite{dataset_yousefi,dataset_widar3,dataset_operanet}. 
				These activities are: (1) Nothing, (2) Walking, (3) Rotation, (4) Jumping, (5) Waving, (6) Lying Down, (7) Picking Up, (8) Sitting Down, (9) Standing Up.
				To build up a comprehensive dataset, we collect data in 3 daily environments: (1) Classroom, (2) Meeting Room, (3) Empty Room.
				In each environment, we specify 5 locations and mark them as A, B, C, D, and E.
				Fig. \ref{figure_layout} describes the layouts of three environments and the interior locations.
				We recruit 6 volunteers to act as users, including 3 females and 3 males, with an average age of 27.33 $\pm$ 0.94, height of 169.00 $\pm$ 6.83 cm, weight of 61.17 $\pm$ 10.04 kg, and Body Mass Index (BMI) of 21.25 $\pm$ 1.78.
				%
				These users are assigned with identity labels (\textit{e.g.}, User 1, User 2, ...) to protect their privacy and to support human identification.
				Regarding the human subject study, we have obtained an approval from institutional ethics committee (IRB) in advance and will provide an ethics statement in Section \ref{section_ethics}.

			%
			\noindent\textbf{Collection Setup.}
				We control the collection of each 3-second sample in three dimensions: (1) users, (2) locations, (3) activities.
				For the user dimension, we organize volunteers into user groups, where each group includes fixed users.
				For the location and activity dimensions, we devise simultaneous scripts for users in each group, so that they know where to stay and what to do independently yet simultaneously.
				User groups and simultaneous scripts are described as follows.

			\noindent\textbf{User Groups.}
				We collect samples by user groups, where each group corresponds to a specific number of users and environment.
				For example, Fig. \ref{figure_act_30_25} shows a sample in the 30th user group, where 3 users (User 1, 5, and 6) are in the classroom.
				Specifically, there are 6 groups for 0 user, 36 groups for 1 user, 18 groups for 2 users, 18 groups for 3 users, 18 groups for 4 users, and 18 groups for 5 users.
				On the other hand, there are 38 groups for each environment.
				Each group contains 99 samples, and we totally obtain 11286 samples for 114 user groups in WiMANS.
				In each user group, we design simultaneous scripts for users to perform activities. 
				
				%
				%
				%
				%
				%
				%
				%
			
\begin{figure}[t]
				\centering
				\hspace{-12mm}
				\includegraphics[width=1.08\linewidth]{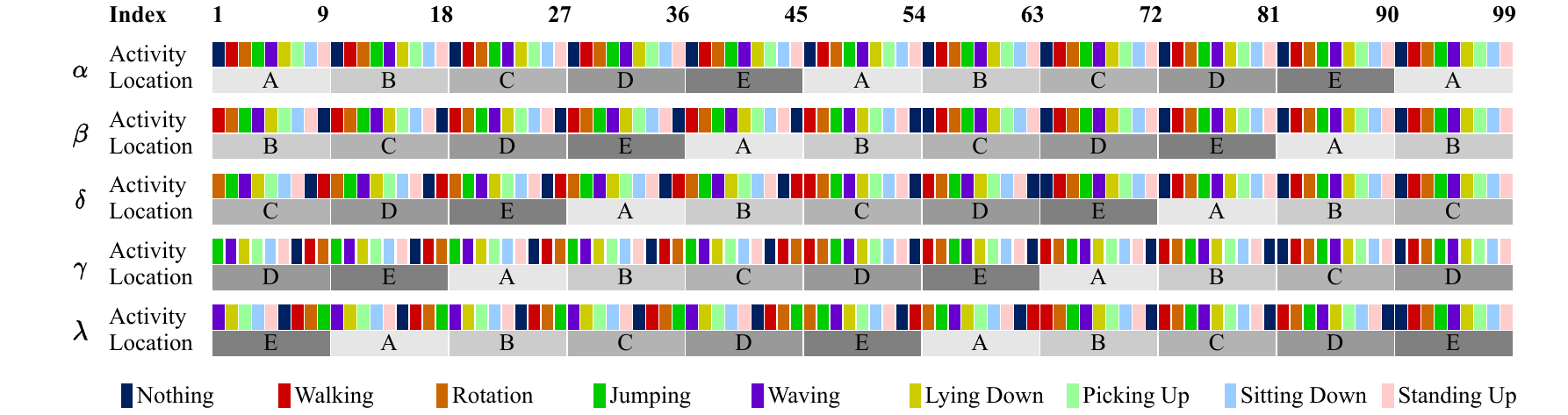}
				\hspace{-10mm}
				\caption{Scripts that instruct users to perform identical/different activities at varying locations independently yet simultaneously.}
				\label{figure_script}
			\end{figure}

			%
			\noindent\textbf{Simultaneous Scripts.}
				%
				%
				In each group, we allocate scripts to fixed users, instructing them to perform activities at different locations independently yet simultaneously.
				To this end, we devise a script set \{$\bm{\alpha}$, $\bm{\beta}$, $\bm{\delta}$, $\bm{\gamma}$, $\bm{\lambda}$\}, as shown in Fig. \ref{figure_script}.
				Each script has 99 indexes, and each index corresponds to an activity at a location.
				Once we announce an index, users with different scripts realize where to stay and what to do.
				%
				%
				%
				For example, in Fig. \ref{figure_act_30_25}, after we announce the index 25, User 1 with $\bm{\alpha}$ performs Picking Up, User 5 with $\bm{\lambda}$ performs Walking, and User 6 with $\bm{\beta}$ performs Sitting Down.
				Note that the scripts are put away before activities to minimize their impacts on data collection.
				Between the collection of each two samples, we first stop the collection of previous sample and then announce a new index to users, after which we start the collection of next sample. 
				Therefore, we can ensure the purity of each sample, not containing nearby activities.
				More details of activities, users, user groups, and simultaneous scripts are provided in Appendix \hyperref[appendix_a]{A}.

			%
			\noindent\textbf{Data Annotation.}
				%
				Since we apply the aforementioned scripts to instruct user activities, we have actually created data annotations before collection.
				%
				We label each sample as ``act\_$<$group$>$\_$<$sample$>$'', where ``$<$group$>$'' is the user group index, and ``$<$sample$>$'' is the sample index.
				For example, Fig. \ref{figure_act_30_25} shows the sample labeled as ``act\_30\_25'', manifesting that it is the 25th sample of the 30th user group.
				%
				We associate each label with the environment, WiFi band, number of users, identities, locations, and activities, to enable different sensing tasks.
				Note that the videos share the same labels with CSI and thereby can act as references to support unexplored tasks (\textit{e.g.}, multi-user pose estimation).

		\subsection{Dataset Statistics}
			\label{subsection_dataset_statistic}

			\begin{figure}[t]
				\centering
				\begin{subfigure}{0.46\linewidth}
					\centering
					\includegraphics[width=\linewidth]{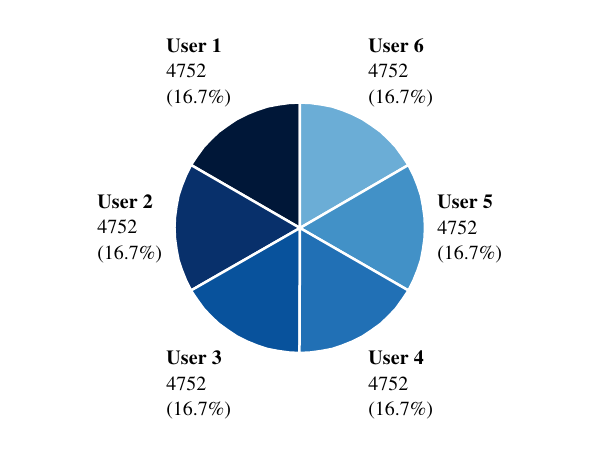}
					\caption{Identity distribution.}
					\label{figure_identity_distribution}
				\end{subfigure}
				\begin{subfigure}{0.46\linewidth}
					\centering
					\includegraphics[width=\linewidth]{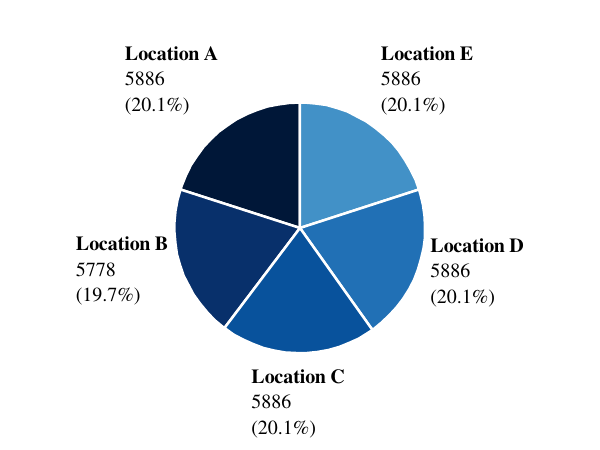}
					\caption{Location distribution.}
					\label{figure_location_distribution}
				\end{subfigure}
				
				\begin{subfigure}{0.46\linewidth}
					\centering
					\includegraphics[width=\linewidth]{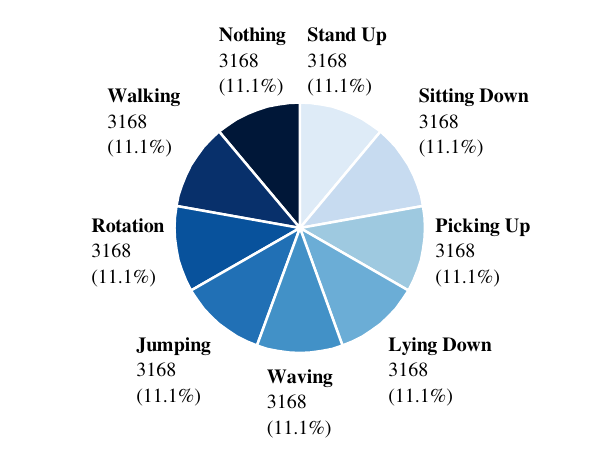}
					\caption{Activity distribution.}
					\label{figure_activity_distribution}
				\end{subfigure}
				\begin{subfigure}{0.46\linewidth}
					\centering
					\includegraphics[width=\linewidth]{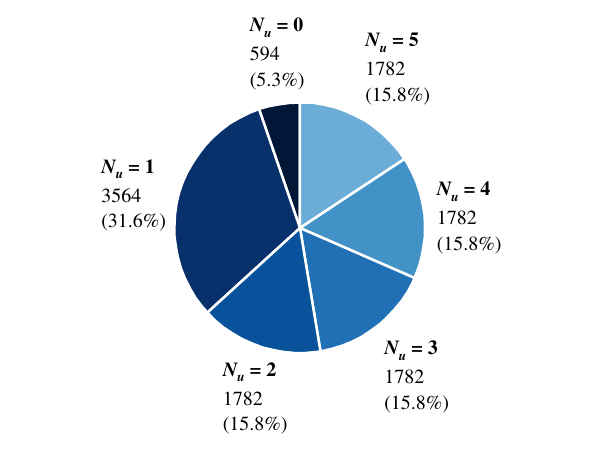}
					\caption{Numbers of users per sample.}
					\label{figure_num_user}
				\end{subfigure}
				\caption{Statistics of WiMANS regarding the distributions of user identities, locations, activities, and numbers of users per sample. In each distribution, all categories have almost equivalent proportions to construct WiMANS as a relatively balanced dataset.}
				\label{figure_statistics}
			\end{figure}

			\noindent\textbf{Data Distribution.}
				%
				WiMANS incorporates varying numbers of users in 11286 samples.
				594 samples (5.3\%) contain 0 user; 3564 samples (31.6\%) contain 1 user; 7128 (63.2\%) samples contain multiple users.
				Among the multi-user samples, there are 1782 (15.8\%) samples for 2 users, 1782 (15.8\%) samples for 3 users, 1782 (15.8\%) samples for 4 users, and 1782 (15.8\%) samples for 5 users.
				Fig. \ref{figure_statistics} presents the statistics of WiMANS. 
				%
				There are 4752 samples for each user, 5778$\sim$5886 samples for each location, and 3168 samples for each activity, empowering models to learn representative features for different sensing tasks.
				Note that the sum of identities/locations/activities is not equal to the total sample number (11286), since each multi-user sample involves more than one identity, location, and activity.

			\noindent\textbf{Packet Loss.}
				%
				Ideally, each CSI sample consists of 3000 time steps, as mentioned in Section \ref{subsection_dataset_construction}. 
				However, packet loss inevitably exists in wireless communication, leading to missing time steps in CSI samples.
				The average packet loss rates are 4.52\% and 2.31\% for the 2.4 GHz band and the 5 GHz band, respectively.
				On average, 2.4 GHz samples have 2864.39 time steps, while 5 GHz samples have 2930.72 time steps.
				2.4 GHz band suffers from more severe packet loss because it is more commonly used and more crowded than 5 GHz band, affected by more environmental noise \cite{dataset_yang,wifi_csi_0}.
				Thus, most existing works \cite{dataset_yousefi,dataset_widar3,dataset_operanet} only apply the 5 GHz band for WiFi-based human sensing.
				However, compared to 5 GHz signals, 2.4 GHz signals can theoretically cover a larger area using longer wavelength to better penetrate and/or bypass obstacles.
				This inspires us to study dual-band augmented human sensing in the future, since WiMANS has gathered CSI of both 2.4 GHz and 5 GHz WiFi bands.
	
	%
	%
	%
	%
	%
	%
	%
	%
	%
	%
	%
	%
	%
	%
	%
	%
	%
	%
	%
	%
	\section{Experiments}
		\label{section_benchmark}

		This section utilizes WiMANS to benchmark the multi-user sensing performance of state-of-the-art WiFi-based and video-based models, with respect to human identification, localization, and activity recognition (HAR).
		We also compare these models in terms of model complexity and time efficiency.
		%
			%
		%
		\subsection{Baselines}
			\label{subsection_baselines}
			\noindent\textbf{WiFi-based models.}
				%
				We evaluate 8 WiFi-based models on WiMANS, including Random Forest based on Short-time Fourier Transform (ST-RF) \cite{dataset_yousefi}, MLP \cite{dataset_ntufi}, LSTM \cite{dataset_yousefi}, CNN-1D \cite{model_cnn_1d}, CNN-2D \cite{model_cnn_2d}, CLSTM \cite{model_clstm}, ABLSTM \cite{model_ablstm}, and THAT \cite{model_that}.
				Particularly, CLSTM \cite{model_clstm} is a hybrid of CNNs and LSTM, outperforming other models in WiFi-based identification.
				For WiFi-based localization, CNN-1D \cite{model_cnn_1d} has demonstrated the best performance by learning human features along the temporal dimension.
				THAT \cite{model_that} equips two-stream Transformer encoders \cite{ml_attention} with multi-scale convolutions and achieves state-of-the-art performance in WiFi-based HAR.
				%
			
			%
			\noindent\textbf{Video-based models.}
				%
				We apply 6 state-of-the-art video classification models for comparison, including ResNet \cite{model_resnet}, S3D \cite{model_s3d}, MViT-v1 \cite{model_mvit_v1}, MViT-v2 \cite{model_mvit_v2}, Swin-T \cite{model_swin}, and Swin-S \cite{model_swin}.
				These models have demonstrated state-of-the-art performance in generic video classification.

		\subsection{Evaluation Metrics}
			\label{subsection_metrics}
			%
				%
				We employ accuracy to measure the recognition performance of multi-user sensing, following the previous works \cite{dataset_yousefi,model_ablstm,model_that}.
				%
				%
				Note that each multi-user sample includes multiple identity/location/activity labels corresponding to different users, respectively.
				For example, Fig. \ref{figure_act_30_25} shows a sample involving three identities: User 1, User 5, and User 6.
				Therefore, we measure the accuracy of recognizing these labels in each sensing task, rather than calculating the accuracy of classifying each sample.
				%
				%
				To evaluate model complexity and time efficiency, we adopt the number of parameters, floating point operations (FLOPs), and recognition throughput as metrics, following the previous works \cite{model_that,dataset_ntufi}.

		\subsection{Implementation Details}
			\label{subsection_implementation}
			\noindent\textbf{Data Preprocessing.}
				We analyze model performance in different environments and accordingly split WiMANS into 3 subsets.
				Each subset is randomly split into a training set (80\%) and a test set (20\%) for evaluation.
				Following the previous works \cite{dataset_yousefi,model_ablstm,model_that}, we calculate and utilize the amplitudes of CSI for the evaluation of WiFi-based models.
				Inputs of all models are resized following the original papers (\textit{e.g.}, 90$\times$3$\times$112$\times$112 for ResNet \cite{model_resnet}).

			\noindent\textbf{Hyperparameters.}
				(1) For WiFi-based models, we leverage publicly accessible implementations as much as possible. 
				Specifically, we initialize WiFi-based models with Xavier \cite{experiment_xavier} and use a fixed learning rate of 10$^{-3}$ to train them for 200 epochs with the batch size of 128.
				(2) For video-based models, we utilize the implementations provided by PyTorch and initialize them with the weights pre-trained by Kinetics-400 \cite{experiment_kinetics}.
				All video-based models are trained for 20 epochs with a fixed learning rate of 10$^{-4}$ and the batch size of 8.

			\noindent\textbf{Evaluation.}
				To recognize multiple labels, we connect each model to a linear layer followed by a sigmoid function for multi-label classification, using a fixed threshold of 0.5 to determine identities, locations, and activities.
				Both WiFi-based and video-based models are optimized by Adam \cite{experiment_adam} on a single Nvidia RTX A5000 GPU.
				%
				%
				We repeat each experiment 10 times with random seeds and report the means and standard deviations of results.
				%
				%
				More implementation details are provided in Appendix \hyperref[appendix_c]{C}. 

		\begin{table}[t]
			\centering
			\tiny
			\caption{Recognition performance of WiFi-based models on WiMANS in terms of accuracy (\%). Models show desirable performance for multi-user identification and localization, while there is still vast room for improvement in activity recognition.}
			\begin{tabular}{l *{3}{c@{\;}c@{\;}c} }
				
				\toprule
				
				\multirow{2}*{\tb{Model}}	& \multicolumn{3}{c}{\tb{Classroom}}	& \multicolumn{3}{c}{\bf{Meeting Room}}	& \multicolumn{3}{c}{\tb{Empty Room}}		\vspace{-0.8mm}\\
				
				\cmidrule(lr){2-4}\cmidrule(lr){5-7}\cmidrule(lr){8-10}
				
				~						& Identity				& Location			& Activity	& Identity	& Location	& Activity	& Identity	& Location	& Activity\\ 
				
				\midrule
				\rowcolor[HTML]{EFEFEF} \multicolumn{10}{c}{2.4 GHz}\vspace{1mm}\\
				ST-RF \cite{dataset_yousefi}& 66.3\std{1.33}		& 62.9\std{0.36}		& 56.8\std{0.14}		& 79.0\std{1.32}		& 65.7\std{0.55}		& 56.8\std{0.09}		& 71.0\std{2.20}		& 61.5\std{0.54}		& 56.7\std{0.09}		\\
				MLP	\cite{dataset_ntufi}	& 65.7\std{0.91}		& 60.5\std{0.47}		& 57.6\std{0.13}		& 75.9\std{0.62}		& 64.2\std{0.47}		& 57.8\std{0.16}		& 75.3\std{0.59}		& 58.8\std{0.26}		& 57.0\std{0.12}		\\
				LSTM \cite{dataset_yousefi}	& 80.7\std{1.43}		& 66.9\std{1.09}		& 59.2\std{0.42}		& 89.5\std{1.11}		& 71.7\std{0.88}		& 59.0\std{0.33}		& 86.7\std{0.99}		& 67.8\std{0.65}		& 57.7\std{0.54}		\\
				CNN-1D \cite{model_cnn_1d}	& 84.8\std{0.80}		& 73.2\std{0.22}		& 59.6\std{0.24}		& 92.9\std{0.26}		& 77.2\std{0.19}		& 58.5\std{0.11}		& 88.1\std{0.46}		& 74.5\std{0.31}		& 58.0\std{0.17}		\\
				CNN-2D \cite{model_cnn_2d}	& 90.0\std{0.66}		& 77.0\std{0.37}		& 59.5\std{0.15}		& \tb{96.4\std{0.81}}	& 82.1\std{0.40}		& 59.0\std{0.12}		& 91.1\std{0.82}		& 79.1\std{0.52}		& 58.2\std{0.26}		\\
				CLSTM \cite{model_clstm}	& 90.2\std{0.70}		& 73.0\std{0.68}		& \tb{61.8\std{0.55}}	& 94.6\std{0.49}		& 77.4\std{0.75}		& 60.9\std{0.39}	& 92.8\std{0.27}		& 71.8\std{0.95}		& \tb{61.0\std{0.40}}	\\
				ABLSTM \cite{model_ablstm}	& 88.8\std{1.36}		& 76.6\std{0.42}		& 61.6\std{0.50}		& 94.8\std{0.55}		& 77.7\std{0.61}		& \tb{60.9\std{0.27}}		& 90.9\std{0.54}		& 76.0\std{0.55}		& 59.7\std{0.15}		\\
				THAT \cite{model_that}		& \tb{90.7\std{0.51}}	& \tb{78.8\std{0.74}}	& 61.0\std{0.32}		& 94.8\std{0.59}		& \tb{82.2\std{0.84}}	& 60.1\std{0.28}		& \tb{93.9\std{0.88}}	& \tb{80.8\std{1.11}}	& 59.7\std{0.29}		\\
				\midrule
				\rowcolor[HTML]{EFEFEF} \multicolumn{10}{c}{5 GHz}\vspace{1mm}\\
				ST-RF \cite{dataset_yousefi}& 89.4\std{1.08}		& 62.7\std{0.57}		& 57.3\std{0.08}		& 95.1\std{0.83}		& 67.4\std{0.65}		& 57.6\std{0.13}		& 81.9\std{1.26}		& 61.9\std{0.56}		& 57.2\std{0.21}		\\
				MLP	\cite{dataset_ntufi}	& 98.6\std{0.13}		& 72.7\std{0.82}		& 58.6\std{0.14}		& 99.5\std{0.11}		& 79.1\std{0.42}		& 57.4\std{0.14}		& 95.1\std{0.58}		& 74.5\std{0.87}		& 59.2\std{0.23}		\\
				LSTM \cite{dataset_yousefi}	& 98.9\std{0.16}		& 76.1\std{0.68}		& 60.6\std{0.28}		& 99.7\std{0.17}		& 81.3\std{0.80}		& 59.5\std{0.23}		& 95.8\std{0.50}		& 77.4\std{0.37}		& 59.8\std{0.36}		\\
				CNN-1D \cite{model_cnn_1d}	& 99.5\std{0.21}		& 82.4\std{0.34}		& 60.6\std{0.22}		& 99.7\std{0.12}		& 87.1\std{0.32}		& 59.0\std{0.16}		& 95.9\std{0.36}		& 84.0\std{0.58}		& 60.2\std{0.14}		\\
				CNN-2D \cite{model_cnn_2d}	& 99.1\std{0.21}		& 82.7\std{0.41}		& 59.8\std{0.30}		& 99.8\std{0.11}		& \tb{88.2\std{0.37}}	& 58.6\std{0.29}		& 95.3\std{0.36}		& \tb{84.4\std{0.40}}	& 59.6\std{0.28}		\\
				CLSTM \cite{model_clstm}	& \tb{99.7\std{0.19}}	& 82.0\std{0.50}		& \tb{64.2\std{0.55}}	& 99.8\std{0.08}		& 88.1\std{0.55}		& \tb{62.1\std{0.36}}	& \tb{97.6\std{0.25}}	& 84.4\std{0.64}		& \tb{64.8\std{0.38}}	\\
				ABLSTM \cite{model_ablstm}	& 99.6\std{0.29}		& 82.5\std{0.67}		& 61.4\std{0.22}		& 99.8\std{0.08}		& 87.4\std{0.25}		& 60.4\std{0.16}		& 96.7\std{0.38}		& 83.8\std{0.53}		& 61.3\std{0.15}		\\
				THAT \cite{model_that}		& 99.2\std{0.19}		& \tb{83.1\std{0.67}}	& 61.8\std{0.29}		& \tb{99.9\std{0.00}}	& 88.1\std{0.59}		& 61.2\std{0.37}		& 97.1\std{0.29}		& 83.7\std{1.05}		& 62.1\std{0.41}		\\
				\midrule
				\rowcolor[HTML]{EFEFEF} \multicolumn{10}{c}{2.4 / 5 GHz}\vspace{1mm}\\
				ST-RF \cite{dataset_yousefi}& 77.9\std{1.05}		& 62.5\std{0.39}		& 57.3\std{0.09}		& 87.7\std{0.91}		& 66.7\std{0.34}		& 57.4\std{0.07}		& 76.0\std{1.06}		& 62.1\std{0.28}		& 57.2\std{0.08}		\\
				MLP	\cite{dataset_ntufi}	& 74.9\std{0.48}		& 66.0\std{0.32}		& 57.9\std{0.13}		& 85.9\std{0.78}		& 69.2\std{0.28}		& 57.6\std{0.07}		& 80.1\std{0.53}		& 66.2\std{0.48}		& 58.4\std{0.08}		\\
				LSTM \cite{dataset_yousefi}	& 86.9\std{0.90}		& 69.5\std{0.79}		& 59.6\std{0.24}		& 93.9\std{0.60}		& 75.2\std{0.46}		& 59.7\std{0.31}		& 90.1\std{0.95}		& 70.4\std{0.27}		& 58.9\std{0.25}		\\
				CNN-1D \cite{model_cnn_1d}	& 92.6\std{0.42}		& 78.3\std{0.25}		& 60.7\std{0.14}		& 96.6\std{0.21}		& 82.8\std{0.25}		& 59.5\std{0.17}		& 93.5\std{0.21}		& 78.5\std{0.21}		& 59.7\std{0.12}		\\
				CNN-2D \cite{model_cnn_2d}	& \tb{94.6\std{0.50}}	& 79.2\std{0.40}		& 59.2\std{0.29}		& \tb{98.0\std{0.35}}	& \tb{83.6\std{0.36}}	& 58.5\std{0.28}		& 94.5\std{0.41}		& \tb{80.1\std{0.51}}	& 58.3\std{0.21}		\\
				CLSTM \cite{model_clstm}	& 93.1\std{0.46}		& 79.9\std{0.48}		& \tb{64.0\std{0.42}}	& 97.4\std{0.28}		& 83.1\std{0.33}		& \tb{62.4\std{0.28}}	& 95.3\std{0.38}		& 78.2\std{0.60}		& \tb{63.5\std{0.36}}	\\
				ABLSTM \cite{model_ablstm}	& 92.4\std{0.39}		& 79.2\std{0.38}		& 62.1\std{0.16}		& 97.8\std{0.17}		& 82.8\std{0.42}		& 60.9\std{0.17}		& 95.0\std{0.34}		& 78.9\std{0.54}		& 61.4\std{0.29}		\\
				THAT \cite{model_that}		& 94.3\std{0.51}		& \tb{80.4\std{0.56}}	& 61.8\std{0.19}		& 97.7\std{0.33}		& 83.3\std{0.45}		& 61.2\std{0.22}		& \tb{95.8\std{0.40}}	& 80.0\std{0.82}		& 62.0\std{0.33}		\\
				\bottomrule
				
			\end{tabular}
			\label{table_result_wifi_csi}
		\end{table}

		\subsection{Results and Analysis}
			\label{subsection_comparison_results}

			\noindent\textbf{WiFi-based Multi-user Sensing.}
				Table \ref{table_result_wifi_csi} presents the multi-user sensing performance of WiFi-based models. 
				(1) Using the 2.4 GHz WiFi band, THAT and CNN-2D are the best performing models for identification, while CNN-2D outperforms other models in localization, and CLSTM achieves the best performance in HAR.
				(2) With the 5 GHz WiFi band, CLSTM and THAT show the highest accuracy in recognizing multiple identities, while THAT and CNN-2D demonstrate their superiority in localization, and CLSTM obtains better results than other models in HAR.
				(3) Exploiting both 2.4 and 5 GHz WiFi bands, CNN-2D and THAT have the best results in identifying and localizing multiple users, while CLSTM yields the highest accuracy for HAR.
				Generally, we can observe that the state-of-the-art models show better performance on identification (90.69$\sim$99.99\%) than localization (78.78$\sim$88.20\%), while having even lower performance on HAR (60.87$\sim$64.82\%).
				This is because HAR is more fine-grained than the other two tasks, causing higher difficulties to learn representative features.
				In contrast, multi-user identification is less fine-grained than localization and thereby results in better performance. 
				Comparing different WiFi bands, using 5 GHz results in higher accuracy than 2.4 GHz, since the 2.4 GHz WiFi band suffers from more noise than 5 GHz, as mentioned in Section \ref{subsection_dataset_statistic}.
				Meanwhile, simply using the samples of two WiFi bands together cannot produce models of better performance.
				These results inspire us to study dual-band augmented multi-user sensing in the future. 

			\begin{figure}[t]
				\centering
				\begin{subfigure}{0.32\linewidth}
					\centering
					\includegraphics[width=\linewidth]{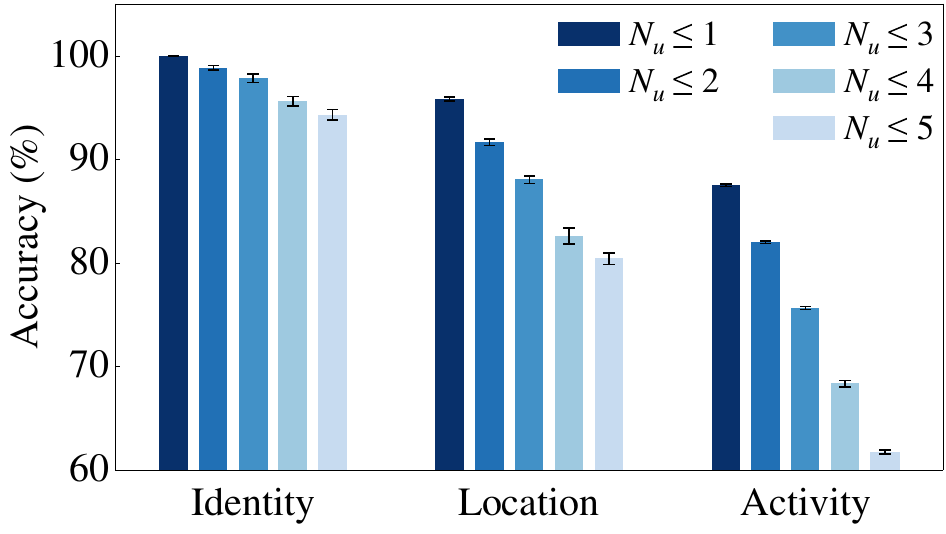}
					\caption{Classroom.}
					\label{figure_nu_classroom}
				\end{subfigure}
				\hfill
				\begin{subfigure}{0.32\linewidth}
					\centering
					\includegraphics[width=\linewidth]{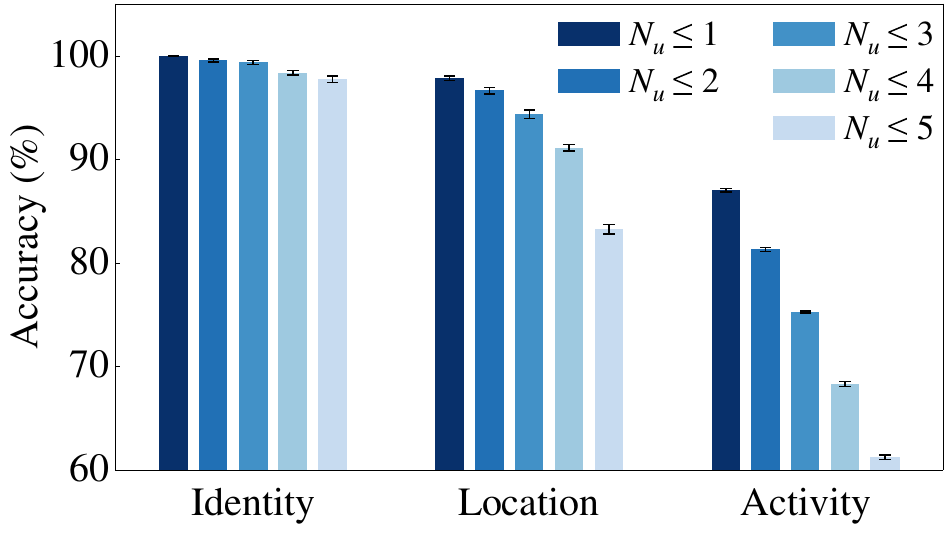}
					\caption{Meeting Room.}
					\label{figure_nu_meeting}
				\end{subfigure}
				\hfill
				\begin{subfigure}{0.32\linewidth}
					\centering
					\includegraphics[width=\linewidth]{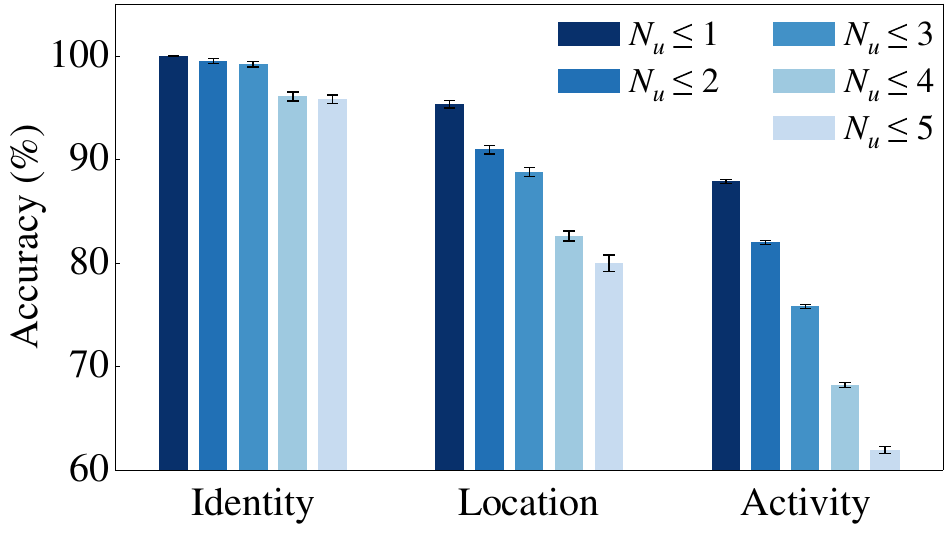}
					\caption{Empty Room.}
					\label{figure_nu_empty}
				\end{subfigure}
				\caption{Recognition performance of THAT \cite{model_that} in terms of accuracy (\%) regarding different numbers of users in each CSI sample ($N_u$). Sensing more users simultaneously results in lower performance. (``$N_u \leq k$'': each sample includes 0 to $k$ users.)}
				\label{figure_number_of_user_performance}
			\end{figure}

			\noindent\textbf{Numbers of Users.}
				Fig. \ref{figure_number_of_user_performance} analyzes the impacts of user numbers on the performance of THAT.
				As expected, sensing more users simultaneously results in lower recognition accuracy.
				(1) In the classroom, when the maximum user number increases from 1 to 5, the accuracy of identification, localization, and HAR decreases by 5.70\%, 15.39\%, and 25.74\%, respectively.
				(2) In the meeting room, sensing up to 5 users simultaneously diminishes the accuracy of three tasks by 2.26\%, 14.59\%, and 25.78\% compared to sensing a single user.
				(3) In the empty room, similar results are evident where the accuracy of three tasks declines by 4.19\%, 15.36\%, and 25.90\% as the maximum user number raises from 1 to 5.
				We can observe that, compared to localization and identification, the performance of HAR is more sensitive to changes in the number of users, because HAR is of finer granularity, and increasing users leads to more occlusion and mutual interference.
				Meanwhile, in varying environments, the number of users exerts distinct impacts on the recognition accuracy, which is worthwhile to further study.
				%
				%
				These results illustrate the usefulness of WiMANS and highlight the challenges of sensing multiple users simultaneously based on WiFi CSI.
			
			%
			\begin{table*}[t]
				\centering
				\tiny
				\caption{Recognition performance of video-based models on WiMANS in terms of accuracy (\%). Video-based models show better performance than WiFi-based models.}
				\begin{tabular}{l *{3}{c@{\;}c@{\;}c} }
					\toprule
					\multirow{2}*{\bf{Model}}	& \multicolumn{3}{c}{\bf{Classroom}}		& \multicolumn{3}{c}{\bf{Meeting Room}}		& \multicolumn{3}{c}{\bf{Empty Room}}	\vspace{-0.8mm}\\
					\cmidrule(lr){2-4}\cmidrule(lr){5-7}\cmidrule(lr){8-10}
					~									& Identity 			& Location 			& Activity 			& Identity 			& Location 			& Activity 			& Identity 			& Location 			& Activity \\ 
					\midrule
					ResNet \cite{model_resnet}			& 99.0\std{0.93} 		& 95.9\std{1.70} 		& 72.3\std{0.86} 		& 99.2\std{0.94} 		& 95.6\std{5.14} 		& 73.6\std{1.13} 		& 96.1\std{1.39} 		& 96.2\std{5.38} 		& 76.5\std{4.74} 	\\
					S3D \cite{model_s3d}				& \tb{99.9\std{0.00}} 	& 99.5\std{0.50} 		& 94.0\std{1.35} 		& 99.9\std{0.04} 		& \tb{99.9\std{0.02}} 	& \tb{98.0\std{0.12}} 	& 99.7\std{0.32} 		& \tb{99.9\std{0.01}} 	& \tb{98.5\std{0.15}} 	\\
					MViT-v1 \cite{model_mvit_v1}		& 99.9\std{0.05} 		& 99.3\std{0.26} 		& 93.7\std{0.44} 		& 99.7\std{0.12} 		& 99.6\std{0.05} 		& 90.7\std{0.11} 		& 99.7\std{0.04} 		& 99.8\std{0.01} 		& 94.0\std{0.10} 	\\
					MViT-v2 \cite{model_mvit_v2}		& 99.8\std{0.06} 		& \tb{99.7\std{0.05}} 	& 95.1\std{0.11} 		& 99.8\std{0.13} 		& 99.8\std{0.04} 		& 92.3\std{0.13} 		& 99.8\std{0.12} 		& 99.9\std{0.01} 		& 95.2\std{0.11}	\\
					Swin-T \cite{model_swin}			& 99.8\std{0.05} 		& 99.6\std{0.05} 		& \tb{96.6\std{0.12}} 	& \tb{99.9\std{0.05}} 	& 99.7\std{0.07} 		& 89.1\std{0.14} 		& \tb{99.9\std{0.04}} 	& 99.8\std{0.01} 		& 94.1\std{0.13}	\\
					Swin-S \cite{model_swin}			& 99.9\std{0.07} 		& 99.6\std{0.06} 		& 95.8\std{0.11} 		& 99.5\std{0.08} 		& 99.4\std{0.05} 		& 90.5\std{0.11} 		& 99.8\std{0.12} 		& 99.8\std{0.02} 		& 96.1\std{0.11}	\\
					\bottomrule
				\end{tabular}
				\label{table_result_video}
			\end{table*}

			\begin{table*}[t]
				\centering
				\tiny
				\caption{Model complexity and time efficiency on WiMANS, where WiFi-based models show their superiority over video-based models. (``Recs/s'': recognitions per second.)}
				\begin{tabular}{l@{\quad}l@{\quad}cccccc}
					
					\toprule
					
					\multirow{2}*{\bf{Data}}	& \multirow{2}*{\bf{Model}}	& \multirow{2}*{\bf{Input Size}}	& \multirow{2}*{\bf{Parameters (M)}}	& \multirow{2}*{\bf{FLOPs (G)}}	& \multicolumn{3}{c}{\bf{Throughput (Recs/s)}}	\vspace{-0.8mm}\\
					
					\cmidrule(lr){6-8}
					
					~	& ~		& ~ 	& ~ 	& \makebox[12mm]{~} 	& \makebox[12mm]{Identity} 		& \makebox[11mm]{Location}		& \makebox[12mm]{Activity} \\ 
					
					\midrule
					
					\multirow{7}{*}{WiFi CSI} 	& MLP \cite{dataset_ntufi}			& 810000									& {209.020}		& \tb{0.418}	& 2918.97		& \tb{3399.19}	& \tb{3385.97}	\\
					~							& LSTM \cite{dataset_yousefi}		& 3000$\times$270							& 1.609			& 0.971			& \tb{3047.69}	& 3045.22		& 3058.85		\\
					~							& CNN-1D \cite{model_cnn_1d}		& 3000$\times$270							& 1.916			& 0.516			& 2685.73		& 2679.96		& 2670.38		\\
					~							& CNN-2D \cite{model_cnn_2d}		& 3000$\times$270							& \tb{0.893}	& 1.691			& 2132.54		& 2183.23		& 2137.92		\\
					~							& CLSTM \cite{model_clstm}			& 3000$\times$270							& 5.391			& 1.791			& 2704.11		& 2707.01		& 2754.48		\\
					~							& ABLSTM \cite{model_ablstm}		& 3000$\times$270							& 4.268			& {3.208}		& 2493.05		& 2614.77		& 2708.44		\\
					~							& THAT \cite{model_that}			& 3000$\times$270							& 4.900			& 1.650			& {1937.48}		& {1971.58}		& {1955.32}		\\
					
					\midrule
					
					\multirow{6}{*}{Video} 	& ResNet \cite{model_resnet}			& 90$\times$3$\times$112$\times$112			& 33.393		& \tb{17.670}	& 50.81			& 51.10			& 51.10			\\
					~						& S3D \cite{model_s3d}					& 90$\times$3$\times$224$\times$224			& \tb{8.342}	& 204.644		& \tb{113.86}	& \tb{120.74}	& \tb{117.29}	\\
					~						& MViT-v1 \cite{model_mvit_v1}			& 90$\times$3$\times$224$\times$224			& 36.632		& {632.698}		& 47.93			& 48.03			& 47.94			\\
					~						& MViT-v2 \cite{model_mvit_v2}			& 90$\times$3$\times$224$\times$224			& 34.559		& 577.742		& 38.24			& 38.17			& 38.16			\\
					~						& Swin-T \cite{model_swin}				& 90$\times$3$\times$224$\times$224			& 28.180		& 268.797		& 44.90			& 44.80			& 44.83			\\
					~						& Swin-S \cite{model_swin}				& 90$\times$3$\times$224$\times$224			& {49.838}		& 518.993		& {27.40}		& {27.43}		& {27.41}			\\
					
					\bottomrule
					
				\end{tabular}
				\label{table_result_complexity}
			\end{table*}

			\noindent\textbf{Video-based Multi-user Sensing.}
				We provide the performance of video-based models in Table \ref{table_result_video} for comparison.
				Overall, video-based models achieve better performance than WiFi-based models in three sensing tasks.
				Such results are expected since video-based analysis has been well studied in computer vision, which also demonstrate the quality of WiMANS, where users have performed distinguishing activities.
				Therefore, we can further exploit the videos in WiMANS as a reference for unexplored tasks in the future.
				Despite the remarkable performance of video-based models, they have much higher complexity and lower efficiency than WiFi-based models, as discussed below.

			\noindent\textbf{Model Complexity and Time Efficiency.}
				Table \ref{table_result_complexity} highlights the superiority of WiFi-based models compared with video-based models in terms of model complexity and time efficiency.
				Except for MLP which is rarely used in practice due to excessive parameters, the maximum number of parameters (5.391 M for CLSTM) in WiFi-based models is 1.55$\times$ less than the minimum number of parameters (8.342 M for S3D) in video-based models.
				Similarly, the maximum FLOPs (3.208 G for ABLSTM) in WiFi-based models is 5.51$\times$ lower than the minimum FLOPs (17.670 G for ResNet) in video-based models.
				Moreover, the throughput of THAT is 16.33$\sim$17.02$\times$ higher than the throughput of S3D.
				These results signify that WiFi-based human sensing can achieve satisfactory performance with higher cost-effectiveness than video-based solutions.
				We further discuss the training time and testing time of models in Appendix \hyperref[appendix_c]{C}.
				%
				%
	
	%
	%
	%
	%
	%
	%
	%
	%
	%
	%
	%
	%
	%
	%
	%
	%
	%
	%
	%
	%
	\section{Discussion}
		\label{section_discussion}
		To the best of our knowledge, WiMANS is the first benchmark dataset for WiFi-based multi-user activity sensing. 
		In this section, we discuss the limitations and future work of WiMANS and provide an ethics statement. 
		%
		
		%
		\subsection{Limitations}
			%
			WiMANS aims to represent the most common multi-user scenarios in daily life.
			Therefore, we discuss the most general settings but not specific activities, conditions, dedicated devices, \textit{etc.}
			Such general settings may lead to some limitations. 
			%
			
			%
			\noindent\textbf{Daily Activities.}
			WiMANS includes common daily activities only, not ethically risky activities (\textit{e.g.}, falling, fighting),
			and the current activities in WiMANS are representative of daily life \cite{dataset_yousefi,dataset_widar3,dataset_operanet}. 
			We will further consider other activities in our future work, given ethical approval to ensure the safety of volunteers.

			\noindent\textbf{Challenging Conditions.}
			WiFi-based human sensing has the potential to solve the issues of obstacles.
			However, we do not intentionally set up obstacles since we use a camera to capture videos, though WiMANS does contain occlusion due to obstacles (\textit{e.g.}, chairs, desks) and indeed people also overlap with each other.
			%

			%
			\noindent\textbf{WiFi Devices.}
			We collect WiFi CSI using a single transmitter-receiver pair equipped with Intel 5300 Network Interface Cards, the most commonly used devices in the previous works \cite{dataset_yousefi,dataset_signfi,dataset_falldefi,dataset_wiar,dataset_widar3,dataset_operanet}.
			More transmitter-receiver pairs may benefit WiFi-based multi-user sensing \cite{related_multi_device},
			while other dedicated devices can also collect CSI using the Atheros CSI tool \cite{dataset_mmfi,related_atheros_csi}, the AX-CSI tool \cite{dataset_sharpax,related_ax_csi}, \textit{etc}.
			%
			%
		
		%
		\subsection{Future Work}
			%
			\noindent\textbf{Multi-user Pose Estimation.}
			Recent works have discussed WiFi-based pose estimation \cite{related_pose_0,related_pose_1,related_pose_2,related_pose_3,related_pose_4,related_pose_5,related_pose_6}, but they have not released their datasets.
			Since WiMANS collects both WiFi CSI and videos, we can extend our dataset by annotating videos with human body joints for multi-user pose estimation.
			
			%
			\noindent\textbf{Dual-band Augmented Sensing.}
				2.4 GHz and 5 GHz WiFi bands have different advantages for human sensing, as mentioned in \ref{subsection_dataset_statistic}.
				We plan to investigate multi-user sensing augmented by dual WiFi bands for better performance.
			
				%
			
			%
			\noindent\textbf{Cross-domain Sensing.}
				Recent studies \cite{survey_2023,related_domain_adaptation} have discussed cross-domain single-user sensing based on WiFi CSI.
				We plan to explore cross-domain multi-user sensing with varying numbers of users, unseen environments/users, \textit{etc}.

		\subsection{Ethics Statement}
			\label{section_ethics}
			%
			Human subject study in WiMANS has been reviewed and approved by institutional ethics committee (IRB).
			%
			%
			Prior to data collection, each subject has been given the details of WiMANS and signed the consent form regarding safety, privacy, and releasing identifiable information.
			The activities in WiMANS are common in daily life (\textit{e.g.}, Walking) to minimize participant risks. 
			We create aliases for all subjects (\textit{e.g.}, User 1) and blur their faces in videos to anonymize personally identifiable information. 
			%
			WiMANS should be used for academic research purposes only.
			We provide further discussions in Appendix \hyperref[appendix_d]{D}.
	
	%
	%
	%
	%
	%
	%
	%
	%
	%
	%
	%
	%
	%
	%
	%
	%
	%
	%
	%
	%
	\section{Conclusion}
		\label{section_conclusion}
		In this paper, we introduce WiMANS, the first WiFi-based human sensing dataset that incorporates multiple users in each CSI sample.
		%
		WiMANS contains 11286 CSI samples of dual WiFi bands and synchronized videos, annotated with user identities, locations, and activities.
		%
		We run benchmark experiments on WiMANS to evaluate the multi-user sensing performance of existing models based on WiFi and videos.
		%
		WiFi-based models show desirable performance for multi-user identification and localization, while there is still vast room for improvement in activity recognition compared with video-based models.
		%
		We make all data, source code, and documents publicly available, anticipating that WiMANS can advance WiFi-based human sensing research from single-user sensing towards more practical multi-user sensing.
		%
		We hope that WiMANS can enable a variety of future works, such as multi-user pose estimation, dual-band augmented sensing, and cross-domain sensing.
		%

	
	\appendix
	\setcounter{table}{0}
	\setcounter{figure}{0}
	\renewcommand{\tablename}{Supplement Table}
	\renewcommand{\figurename}{Supplement Fig.}
	\section*{Appendix}
	%
	%
	%
	%
	%
	%
	%
	%
	%
	%
	%
	%
	%
	%
	%
	%
	%
	%
	%
	%
	\section{Data Collection}
	\label{appendix_a}
	Herein, we provide more details of data collection in WiMANS, including the description of activities (\ref{subsection_description_activity}), user statistics (\ref{subsection_user_statistics}), and the details of 114 user groups (\ref{subsection_user_group}).
	\subsection{Description of Activities}
	\label{subsection_description_activity}
\begin{figure*}[htbp]
		\centering
		\begin{subfigure}{\linewidth}
			\centering
			\includegraphics[width=0.95\linewidth]{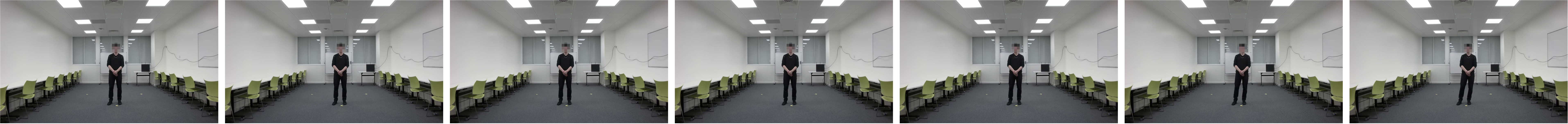}
			\caption{Nothing: A user is standing still and doing nothing at a designated location.}
			\label{figure_act_74_18}
		\end{subfigure}
		
		\begin{subfigure}{\linewidth}
			\centering
			\includegraphics[width=0.95\linewidth]{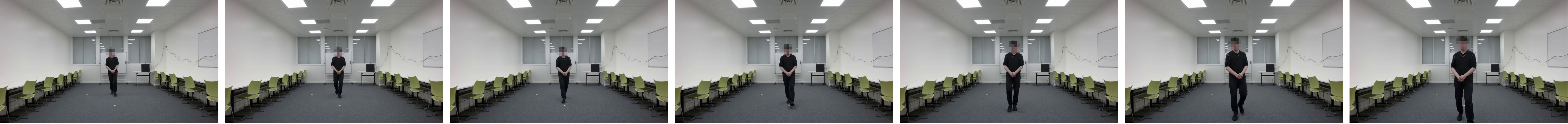}
			\caption{Walking: A user is walking through a designated location.}
			\label{figure_act_74_10}
		\end{subfigure}
		
		\begin{subfigure}{\linewidth}
			\centering
			\includegraphics[width=0.95\linewidth]{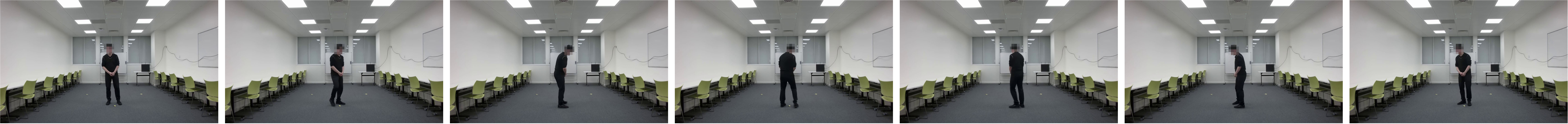}
			\caption{Rotation: A user is rotating at a designated location.}
			\label{figure_act_74_11}
		\end{subfigure}
		
		\begin{subfigure}{\linewidth}
			\centering
			\includegraphics[width=0.95\linewidth]{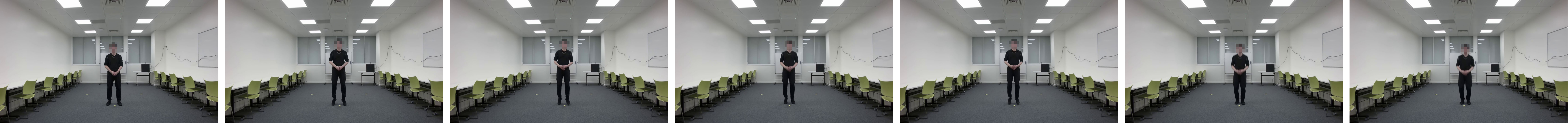}
			\caption{Jumping: A user is jumping at a designated location.}
			\label{figure_act_74_12}
		\end{subfigure}
		
		\begin{subfigure}{\linewidth}
			\centering
			\includegraphics[width=0.95\linewidth]{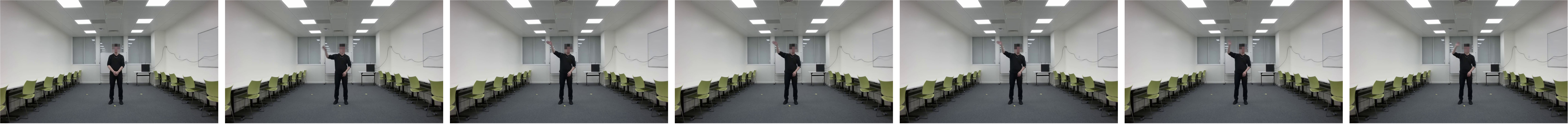}
			\caption{Waving: A user is waving hand(s) at a designated location.}
			\label{figure_act_74_13}
		\end{subfigure}
		
		\begin{subfigure}{\linewidth}
			\centering
			\includegraphics[width=0.95\linewidth]{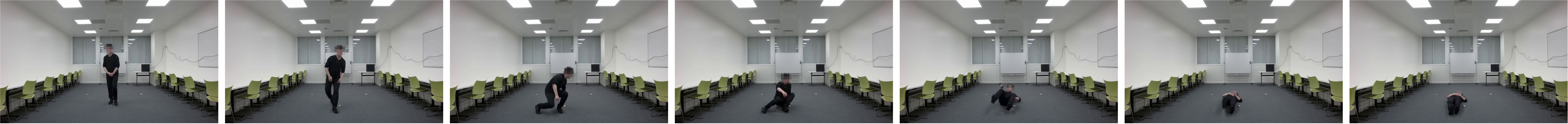}
			\caption{Lying Down: A user is lying down on the floor, before which this user was standing still.}
			\label{figure_act_74_14}
		\end{subfigure}
		
		\begin{subfigure}{\linewidth}
			\centering
			\includegraphics[width=0.95\linewidth]{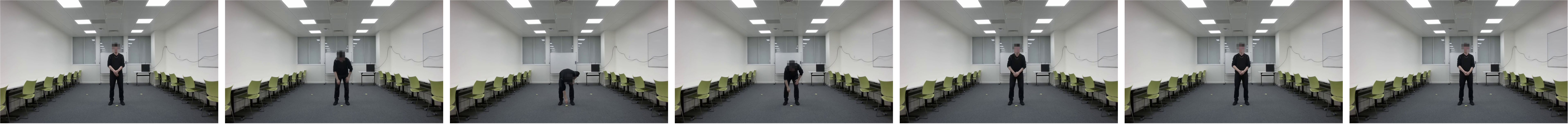}
			\caption{Picking Up: A user is picking something up from the floor.}
			\label{figure_act_74_15}
		\end{subfigure}
		
		\begin{subfigure}{\linewidth}
			\centering
			\includegraphics[width=0.95\linewidth]{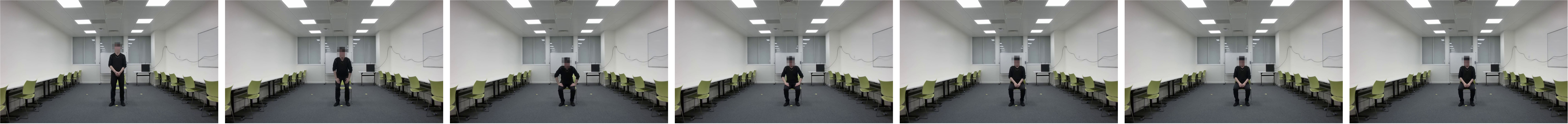}
			\caption{Sitting Down: A user is sitting down on a chair, before which this user was standing still.}
			\label{figure_act_74_16}
		\end{subfigure}
		
		\begin{subfigure}{\linewidth}
			\centering
			\includegraphics[width=0.95\linewidth]{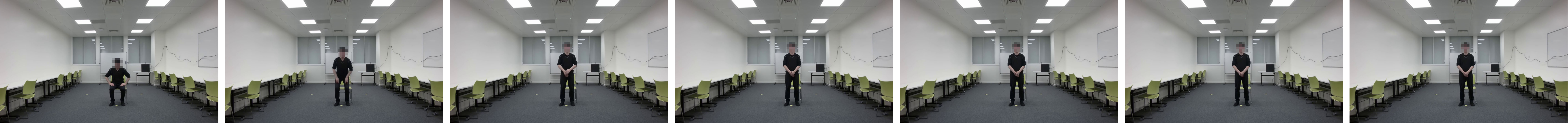}
			\caption{Standing Up: A user is standing up from a chair, before which this user was sitting on the chair.}
			\label{figure_act_74_17}
		\end{subfigure}
		\caption{Descriptions and examples of 9 daily activities in WiMANS.}
		\label{figure_activity}
	\end{figure*}

	WiMANS consists of 9 daily activities: (a) Nothing, (b) Walking, (c) Rotation, (d) Jumping, (e) Waving, (f) Lying Down, (g) Picking Up, (h) Sitting Down, (i) Standing Up.
	Supplement Fig. \ref{figure_activity} presents the descriptions and examples of these activities.

	\subsection{User Statistics}
	\label{subsection_user_statistics}
	We recruit 6 volunteers to act as users in WiMANS, the statistics of users are presented in Supplement Table \ref{table_users}.
	We understand that the users are within a similar age group, and we plan to include more users from a broader age range in the extension of our dataset.
	\begin{table*}[b]
		\centering
		\tiny
		\caption{Statistics of 6 users in WiMANS.}
		\begin{tabular}{cccccc}
			\toprule
			\tb{User Identity}	& \tb{Gender}	& \tb{Age}	& \tb{Weight (kg)}	& \tb{Height (cm)}	& \tb{Body Mass Index (BMI)}		\\
			\midrule
			User 1				& Female		& 27		& 53				& 166				& 19.23						\\
			User 2				& Male			& 29		& 70				& 175				& 22.86						\\
			User 3				& Female		& 26		& 54				& 162				& 20.58						\\
			User 4				& Female		& 28		& 50				& 161				& 19.29						\\
			User 5				& Male			& 27		& 62				& 170				& 21.45						\\
			User 6				& Male			& 27		& 78				& 180				& 24.07						\\
			\bottomrule
		\end{tabular}
		\label{table_users}
	\end{table*}

	\subsection{User Groups}
	\label{subsection_user_group}
	In WiMANS, we organize volunteers into user groups which correspond to different numbers of user and environments.
	In each group, each user holds different scripts (\textit{i.e.}, $\bm{\alpha}$, $\bm{\beta}$, $\bm{\delta}$, $\bm{\gamma}$, and $\bm{\lambda}$) to perform activities independently yet simultaneously.
	Supplement Tables \ref{table_group_1_18} $\sim$ \ref{table_group_200_206} present the details of 114 user groups in WiMANS.

	\begin{table*}[htbp]
		\centering
		\tiny
		\caption{User group 1$\sim$18 in the classroom using the 2.4 GHz WiFi band.}
		\begin{tabular}{ccccccccc}
			\toprule
			\tb{Group}	& \tb{Environment}	& \tb{\# of Users}	& \tb{User 1}	& \tb{User 2}	& \tb{User 3}	& \tb{User 4}	& \tb{User 5}	& \tb{User 6}	\\
			\midrule
			1			& ~								& ~ 				& $\bm{\alpha}$	& ~				& ~				& ~				& ~				& ~				\\
			\cmidrule{4-9}
			2			& ~								& ~					& ~				& $\bm{\beta}$	& ~				& ~				& ~				& ~				\\
			\cmidrule{4-9}
			3			& ~								& \multirow{2}{*}{1}& ~				& ~				& $\bm{\delta}$	& ~				& ~				& ~				\\
			\cmidrule{4-9}
			4			& ~								& ~					& ~				& ~				& ~				& $\bm{\gamma}$	& ~				& ~				\\
			\cmidrule{4-9}
			5			& ~								& ~					& ~				& ~				& ~				& ~				& $\bm{\lambda}$& ~				\\
			\cmidrule{4-9}
			6			& ~								& ~					& ~				& ~				& ~				& ~				& ~				& $\bm{\alpha}$	\\
			\cmidrule{3-9}
			7			& ~								& ~					& $\bm{\alpha}$	& $\bm{\beta}$	& ~				& ~				& ~				& ~				\\
			\cmidrule{4-9}
			8			& ~								& 2					& ~				& ~				& $\bm{\delta}$	& $\bm{\gamma}$	& ~				& ~				\\
			\cmidrule{4-9}
			9			& \multirow{2}{*}{Classroom}	& ~					& ~				& ~				& ~				& ~				& $\bm{\lambda}$& $\bm{\alpha}$	\\
			\cmidrule{3-9}
			10			& ~								& ~					& $\bm{\alpha}$	& $\bm{\beta}$	& $\bm{\delta}$	& ~				& ~				& ~				\\
			\cmidrule{4-9}
			11			& ~								& 3					& ~				& $\bm{\beta}$	& ~				& $\bm{\gamma}$	& ~				& $\bm{\alpha}$	\\
			\cmidrule{4-9}
			12			& ~								& ~					& $\bm{\alpha}$	& ~				& ~				& ~				& $\bm{\lambda}$& $\bm{\beta}$	\\
			\cmidrule{3-9}
			13			& ~								& ~					& $\bm{\alpha}$	& $\bm{\beta}$	& $\bm{\delta}$	& $\bm{\gamma}$	& ~				& ~				\\
			\cmidrule{4-9}
			14			& ~								& 4					& ~				& $\bm{\beta}$	& ~				& $\bm{\gamma}$	& $\bm{\lambda}$& $\bm{\alpha}$	\\
			\cmidrule{4-9}
			15			& ~								& ~					& $\bm{\alpha}$	& ~				& $\bm{\delta}$	& ~				& $\bm{\lambda}$& $\bm{\beta}$	\\
			\cmidrule{3-9}
			16			& ~								& ~					& $\bm{\alpha}$	& $\bm{\beta}$	& $\bm{\delta}$	& $\bm{\gamma}$	& $\bm{\lambda}$& ~				\\
			\cmidrule{4-9}
			17			& ~								& 5					& ~				& $\bm{\beta}$	& $\bm{\delta}$	& $\bm{\gamma}$	& $\bm{\lambda}$& $\bm{\alpha}$	\\
			\cmidrule{4-9}
			18			& ~								& ~					& $\bm{\alpha}$	& ~				& $\bm{\delta}$	& $\bm{\gamma}$	& $\bm{\lambda}$& $\bm{\beta}$	\\
			\bottomrule
		\end{tabular}
		\label{table_group_1_18}
	\end{table*}

	\begin{table*}[htbp]
		\centering
		\tiny
		\caption{User group 19$\sim$36 in the classroom using the 5 GHz WiFi band.}
		\begin{tabular}{ccccccccc}
			\toprule
			\tb{Group}	& \tb{Environment}	& \tb{\# of Users}	& \tb{User 1}	& \tb{User 2}	& \tb{User 3}	& \tb{User 4}	& \tb{User 5}	& \tb{User 6}	\\
			\midrule
			19			& ~								& ~ 				& $\bm{\alpha}$	& ~				& ~				& ~				& ~				& ~				\\
			\cmidrule{4-9}
			20			& ~								& ~					& ~				& $\bm{\beta}$	& ~				& ~				& ~				& ~				\\
			\cmidrule{4-9}
			21			& ~								& \multirow{2}{*}{1}& ~				& ~				& $\bm{\delta}$	& ~				& ~				& ~				\\
			\cmidrule{4-9}
			22			& ~								& ~					& ~				& ~				& ~				& $\bm{\gamma}$	& ~				& ~				\\
			\cmidrule{4-9}
			23			& ~								& ~					& ~				& ~				& ~				& ~				& $\bm{\lambda}$& ~				\\
			\cmidrule{4-9}
			24			& ~								& ~					& ~				& ~				& ~				& ~				& ~				& $\bm{\alpha}$	\\
			\cmidrule{3-9}
			25			& ~								& ~					& $\bm{\alpha}$	& $\bm{\beta}$	& ~				& ~				& ~				& ~				\\
			\cmidrule{4-9}
			26			& ~								& 2					& ~				& ~				& $\bm{\delta}$	& $\bm{\gamma}$	& ~				& ~				\\
			\cmidrule{4-9}
			27			& \multirow{2}{*}{Classroom}	& ~					& ~				& ~				& ~				& ~				& $\bm{\lambda}$& $\bm{\alpha}$	\\
			\cmidrule{3-9}
			28			& ~								& ~					& $\bm{\alpha}$	& $\bm{\beta}$	& $\bm{\delta}$	& ~				& ~				& ~				\\
			\cmidrule{4-9}
			29			& ~								& 3					& ~				& $\bm{\beta}$	& ~				& $\bm{\gamma}$	& ~				& $\bm{\alpha}$	\\
			\cmidrule{4-9}
			30			& ~								& ~					& $\bm{\alpha}$	& ~				& ~				& ~				& $\bm{\lambda}$& $\bm{\beta}$	\\
			\cmidrule{3-9}
			31			& ~								& ~					& $\bm{\alpha}$	& $\bm{\beta}$	& $\bm{\delta}$	& $\bm{\gamma}$	& ~				& ~				\\
			\cmidrule{4-9}
			32			& ~								& 4					& ~				& $\bm{\beta}$	& ~				& $\bm{\gamma}$	& $\bm{\lambda}$& $\bm{\alpha}$	\\
			\cmidrule{4-9}
			33			& ~								& ~					& $\bm{\alpha}$	& ~				& $\bm{\delta}$	& ~				& $\bm{\lambda}$& $\bm{\beta}$	\\
			\cmidrule{3-9}
			34			& ~								& ~					& $\bm{\alpha}$	& $\bm{\beta}$	& $\bm{\delta}$	& $\bm{\gamma}$	& $\bm{\lambda}$& ~				\\
			\cmidrule{4-9}
			35			& ~								& 5					& ~				& $\bm{\beta}$	& $\bm{\delta}$	& $\bm{\gamma}$	& $\bm{\lambda}$& $\bm{\alpha}$	\\
			\cmidrule{4-9}
			36			& ~								& ~					& $\bm{\alpha}$	& ~				& $\bm{\delta}$	& $\bm{\gamma}$	& $\bm{\lambda}$& $\bm{\beta}$	\\
			\bottomrule
		\end{tabular}
		\label{table_group_19_36}
	\end{table*}

	\begin{table*}[htbp]
		\centering
		\tiny
		\caption{User group 37$\sim$54 in the meeting room using the 2.4 GHz WiFi band.}
		\begin{tabular}{ccccccccc}
			\toprule
			\tb{Group}	& \tb{Environment}	& \tb{\# of Users}	& \tb{User 1}	& \tb{User 2}	& \tb{User 3}	& \tb{User 4}	& \tb{User 5}	& \tb{User 6}	\\
			\midrule
			37			& ~								& ~ 				& $\bm{\alpha}$	& ~				& ~				& ~				& ~				& ~				\\
			\cmidrule{4-9}
			38			& ~								& ~					& ~				& $\bm{\beta}$	& ~				& ~				& ~				& ~				\\
			\cmidrule{4-9}
			39			& ~								& \multirow{2}{*}{1}& ~				& ~				& $\bm{\delta}$	& ~				& ~				& ~				\\
			\cmidrule{4-9}
			40			& ~								& ~					& ~				& ~				& ~				& $\bm{\gamma}$	& ~				& ~				\\
			\cmidrule{4-9}
			41			& ~								& ~					& ~				& ~				& ~				& ~				& $\bm{\lambda}$& ~				\\
			\cmidrule{4-9}
			42			& ~								& ~					& ~				& ~				& ~				& ~				& ~				& $\bm{\alpha}$	\\
			\cmidrule{3-9}
			43			& ~								& ~					& $\bm{\alpha}$	& $\bm{\beta}$	& ~				& ~				& ~				& ~				\\
			\cmidrule{4-9}
			44			& ~								& 2					& ~				& ~				& $\bm{\delta}$	& $\bm{\gamma}$	& ~				& ~				\\
			\cmidrule{4-9}
			45			& \multirow{2}{*}{Meeting Room}	& ~					& ~				& ~				& ~				& ~				& $\bm{\lambda}$& $\bm{\alpha}$	\\
			\cmidrule{3-9}
			46			& ~								& ~					& $\bm{\alpha}$	& $\bm{\beta}$	& $\bm{\delta}$	& ~				& ~				& ~				\\
			\cmidrule{4-9}
			47			& ~								& 3					& ~				& $\bm{\beta}$	& ~				& $\bm{\gamma}$	& ~				& $\bm{\alpha}$	\\
			\cmidrule{4-9}
			48			& ~								& ~					& $\bm{\alpha}$	& ~				& ~				& ~				& $\bm{\lambda}$& $\bm{\beta}$	\\
			\cmidrule{3-9}
			49			& ~								& ~					& $\bm{\alpha}$	& $\bm{\beta}$	& $\bm{\delta}$	& $\bm{\gamma}$	& ~				& ~				\\
			\cmidrule{4-9}
			50			& ~								& 4					& ~				& $\bm{\beta}$	& ~				& $\bm{\gamma}$	& $\bm{\lambda}$& $\bm{\alpha}$	\\
			\cmidrule{4-9}
			51			& ~								& ~					& $\bm{\alpha}$	& ~				& $\bm{\delta}$	& ~				& $\bm{\lambda}$& $\bm{\beta}$	\\
			\cmidrule{3-9}
			52			& ~								& ~					& $\bm{\alpha}$	& $\bm{\beta}$	& $\bm{\delta}$	& $\bm{\gamma}$	& $\bm{\lambda}$& ~				\\
			\cmidrule{4-9}
			53			& ~								& 5					& ~				& $\bm{\beta}$	& $\bm{\delta}$	& $\bm{\gamma}$	& $\bm{\lambda}$& $\bm{\alpha}$	\\
			\cmidrule{4-9}
			54			& ~								& ~					& $\bm{\alpha}$	& ~				& $\bm{\delta}$	& $\bm{\gamma}$	& $\bm{\lambda}$& $\bm{\beta}$	\\
			\bottomrule
		\end{tabular}
		\label{table_group_37_54}
	\end{table*}

	\begin{table*}[htbp]
		\centering
		\tiny
		\caption{User group 55$\sim$72 in the meeting room using the 5 GHz WiFi band.}
		\begin{tabular}{ccccccccc}
			\toprule
			\tb{Group}	& \tb{Environment}	& \tb{\# of Users}	& \tb{User 1}	& \tb{User 2}	& \tb{User 3}	& \tb{User 4}	& \tb{User 5}	& \tb{User 6}	\\
			\midrule
			55			& ~								& ~ 				& $\bm{\alpha}$	& ~				& ~				& ~				& ~				& ~				\\
			\cmidrule{4-9}
			56			& ~								& ~					& ~				& $\bm{\beta}$	& ~				& ~				& ~				& ~				\\
			\cmidrule{4-9}
			57			& ~								& \multirow{2}{*}{1}& ~				& ~				& $\bm{\delta}$	& ~				& ~				& ~				\\
			\cmidrule{4-9}
			58			& ~								& ~					& ~				& ~				& ~				& $\bm{\gamma}$	& ~				& ~				\\
			\cmidrule{4-9}
			59			& ~								& ~					& ~				& ~				& ~				& ~				& $\bm{\lambda}$& ~				\\
			\cmidrule{4-9}
			60			& ~								& ~					& ~				& ~				& ~				& ~				& ~				& $\bm{\alpha}$	\\
			\cmidrule{3-9}
			61			& ~								& ~					& $\bm{\alpha}$	& $\bm{\beta}$	& ~				& ~				& ~				& ~				\\
			\cmidrule{4-9}
			62			& ~								& 2					& ~				& ~				& $\bm{\delta}$	& $\bm{\gamma}$	& ~				& ~				\\
			\cmidrule{4-9}
			63			& \multirow{2}{*}{Meeting Room}	& ~					& ~				& ~				& ~				& ~				& $\bm{\lambda}$& $\bm{\alpha}$	\\
			\cmidrule{3-9}
			64			& ~								& ~					& $\bm{\alpha}$	& $\bm{\beta}$	& $\bm{\delta}$	& ~				& ~				& ~				\\
			\cmidrule{4-9}
			65			& ~								& 3					& ~				& $\bm{\beta}$	& ~				& $\bm{\gamma}$	& ~				& $\bm{\alpha}$	\\
			\cmidrule{4-9}
			66			& ~								& ~					& $\bm{\alpha}$	& ~				& ~				& ~				& $\bm{\lambda}$& $\bm{\beta}$	\\
			\cmidrule{3-9}
			67			& ~								& ~					& $\bm{\alpha}$	& $\bm{\beta}$	& $\bm{\delta}$	& $\bm{\gamma}$	& ~				& ~				\\
			\cmidrule{4-9}
			68			& ~								& 4					& ~				& $\bm{\beta}$	& ~				& $\bm{\gamma}$	& $\bm{\lambda}$& $\bm{\alpha}$	\\
			\cmidrule{4-9}
			69			& ~								& ~					& $\bm{\alpha}$	& ~				& $\bm{\delta}$	& ~				& $\bm{\lambda}$& $\bm{\beta}$	\\
			\cmidrule{3-9}
			70			& ~								& ~					& $\bm{\alpha}$	& $\bm{\beta}$	& $\bm{\delta}$	& $\bm{\gamma}$	& $\bm{\lambda}$& ~				\\
			\cmidrule{4-9}
			71			& ~								& 5					& ~				& $\bm{\beta}$	& $\bm{\delta}$	& $\bm{\gamma}$	& $\bm{\lambda}$& $\bm{\alpha}$	\\
			\cmidrule{4-9}
			72			& ~								& ~					& $\bm{\alpha}$	& ~				& $\bm{\delta}$	& $\bm{\gamma}$	& $\bm{\lambda}$& $\bm{\beta}$	\\
			\bottomrule
		\end{tabular}
		\label{table_group_55_72}
	\end{table*}

	\begin{table*}[htbp]
		\centering
		\tiny
		\caption{User group 73$\sim$90 in the empty room using the 2.4 GHz WiFi band.}
		\begin{tabular}{ccccccccc}
			\toprule
			\tb{Group}	& \tb{Environment}	& \tb{\# of Users}	& \tb{User 1}	& \tb{User 2}	& \tb{User 3}	& \tb{User 4}	& \tb{User 5}	& \tb{User 6}	\\
			\midrule
			73			& ~								& ~ 				& $\bm{\alpha}$	& ~				& ~				& ~				& ~				& ~				\\
			\cmidrule{4-9}
			74			& ~								& ~					& ~				& $\bm{\beta}$	& ~				& ~				& ~				& ~				\\
			\cmidrule{4-9}
			75			& ~								& \multirow{2}{*}{1}& ~				& ~				& $\bm{\delta}$	& ~				& ~				& ~				\\
			\cmidrule{4-9}
			76			& ~								& ~					& ~				& ~				& ~				& $\bm{\gamma}$	& ~				& ~				\\
			\cmidrule{4-9}
			77			& ~								& ~					& ~				& ~				& ~				& ~				& $\bm{\lambda}$& ~				\\
			\cmidrule{4-9}
			78			& ~								& ~					& ~				& ~				& ~				& ~				& ~				& $\bm{\alpha}$	\\
			\cmidrule{3-9}
			79			& ~								& ~					& $\bm{\alpha}$	& $\bm{\beta}$	& ~				& ~				& ~				& ~				\\
			\cmidrule{4-9}
			80			& ~								& 2					& ~				& ~				& $\bm{\delta}$	& $\bm{\gamma}$	& ~				& ~				\\
			\cmidrule{4-9}
			81			& \multirow{2}{*}{Empty Room}	& ~					& ~				& ~				& ~				& ~				& $\bm{\lambda}$& $\bm{\alpha}$	\\
			\cmidrule{3-9}
			82			& ~								& ~					& $\bm{\alpha}$	& $\bm{\beta}$	& $\bm{\delta}$	& ~				& ~				& ~				\\
			\cmidrule{4-9}
			83			& ~								& 3					& ~				& $\bm{\beta}$	& ~				& $\bm{\gamma}$	& ~				& $\bm{\alpha}$	\\
			\cmidrule{4-9}
			84			& ~								& ~					& $\bm{\alpha}$	& ~				& ~				& ~				& $\bm{\lambda}$& $\bm{\beta}$	\\
			\cmidrule{3-9}
			85			& ~								& ~					& $\bm{\alpha}$	& $\bm{\beta}$	& $\bm{\delta}$	& $\bm{\gamma}$	& ~				& ~				\\
			\cmidrule{4-9}
			86			& ~								& 4					& ~				& $\bm{\beta}$	& ~				& $\bm{\gamma}$	& $\bm{\lambda}$& $\bm{\alpha}$	\\
			\cmidrule{4-9}
			87			& ~								& ~					& $\bm{\alpha}$	& ~				& $\bm{\delta}$	& ~				& $\bm{\lambda}$& $\bm{\beta}$	\\
			\cmidrule{3-9}
			88			& ~								& ~					& $\bm{\alpha}$	& $\bm{\beta}$	& $\bm{\delta}$	& $\bm{\gamma}$	& $\bm{\lambda}$& ~				\\
			\cmidrule{4-9}
			89			& ~								& 5					& ~				& $\bm{\beta}$	& $\bm{\delta}$	& $\bm{\gamma}$	& $\bm{\lambda}$& $\bm{\alpha}$	\\
			\cmidrule{4-9}
			90			& ~								& ~					& $\bm{\alpha}$	& ~				& $\bm{\delta}$	& $\bm{\gamma}$	& $\bm{\lambda}$& $\bm{\beta}$	\\
			\bottomrule
		\end{tabular}
		\label{table_group_73_90}
	\end{table*}

	\begin{table*}[htbp]
		\centering
		\tiny
		\caption{User group 91$\sim$108 in the empty room using the 5 GHz WiFi band.}
		\begin{tabular}{ccccccccc}
			\toprule
			\tb{Group}	& \tb{Environment}	& \tb{\# of Users}	& \tb{User 1}	& \tb{User 2}	& \tb{User 3}	& \tb{User 4}	& \tb{User 5}	& \tb{User 6}	\\
			\midrule
			91			& ~								& ~ 				& $\bm{\alpha}$	& ~				& ~				& ~				& ~				& ~				\\
			\cmidrule{4-9}
			92			& ~								& ~					& ~				& $\bm{\beta}$	& ~				& ~				& ~				& ~				\\
			\cmidrule{4-9}
			93			& ~								& \multirow{2}{*}{1}& ~				& ~				& $\bm{\delta}$	& ~				& ~				& ~				\\
			\cmidrule{4-9}
			94			& ~								& ~					& ~				& ~				& ~				& $\bm{\gamma}$	& ~				& ~				\\
			\cmidrule{4-9}
			95			& ~								& ~					& ~				& ~				& ~				& ~				& $\bm{\lambda}$& ~				\\
			\cmidrule{4-9}
			96			& ~								& ~					& ~				& ~				& ~				& ~				& ~				& $\bm{\alpha}$	\\
			\cmidrule{3-9}
			97			& ~								& ~					& $\bm{\alpha}$	& $\bm{\beta}$	& ~				& ~				& ~				& ~				\\
			\cmidrule{4-9}
			98			& ~								& 2					& ~				& ~				& $\bm{\delta}$	& $\bm{\gamma}$	& ~				& ~				\\
			\cmidrule{4-9}
			99			& \multirow{2}{*}{Empty Room}	& ~					& ~				& ~				& ~				& ~				& $\bm{\lambda}$& $\bm{\alpha}$	\\
			\cmidrule{3-9}
			100			& ~								& ~					& $\bm{\alpha}$	& $\bm{\beta}$	& $\bm{\delta}$	& ~				& ~				& ~				\\
			\cmidrule{4-9}
			101			& ~								& 3					& ~				& $\bm{\beta}$	& ~				& $\bm{\gamma}$	& ~				& $\bm{\alpha}$	\\
			\cmidrule{4-9}
			102			& ~								& ~					& $\bm{\alpha}$	& ~				& ~				& ~				& $\bm{\lambda}$& $\bm{\beta}$	\\
			\cmidrule{3-9}
			103			& ~								& ~					& $\bm{\alpha}$	& $\bm{\beta}$	& $\bm{\delta}$	& $\bm{\gamma}$	& ~				& ~				\\
			\cmidrule{4-9}
			104			& ~								& 4					& ~				& $\bm{\beta}$	& ~				& $\bm{\gamma}$	& $\bm{\lambda}$& $\bm{\alpha}$	\\
			\cmidrule{4-9}
			105			& ~								& ~					& $\bm{\alpha}$	& ~				& $\bm{\delta}$	& ~				& $\bm{\lambda}$& $\bm{\beta}$	\\
			\cmidrule{3-9}
			106			& ~								& ~					& $\bm{\alpha}$	& $\bm{\beta}$	& $\bm{\delta}$	& $\bm{\gamma}$	& $\bm{\lambda}$& ~				\\
			\cmidrule{4-9}
			107		& ~								& 5					& ~				& $\bm{\beta}$	& $\bm{\delta}$	& $\bm{\gamma}$	& $\bm{\lambda}$& $\bm{\alpha}$	\\
			\cmidrule{4-9}
			108			& ~								& ~					& $\bm{\alpha}$	& ~				& $\bm{\delta}$	& $\bm{\gamma}$	& $\bm{\lambda}$& $\bm{\beta}$	\\
			\bottomrule
		\end{tabular}
		\label{table_group_91_108}
	\end{table*}

	\begin{table*}[htbp]
		\centering
		\tiny
		\caption{User group 201$\sim$206 with 0 user.}
		\begin{tabular}{cccc}
			\toprule
			\tb{Group}	& \tb{Environment}				& \tb{\# of Users}	& \tb{WiFi Band (GHz)}	\\
			\midrule
			201			& \multirow{2}{*}{Classroom}	& \multirow{2}{*}{0}& 2.4					\\
			\cmidrule{4-4}
			202			& ~								& ~					& 5						\\
			\cmidrule{2-4}
			203			& \multirow{2}{*}{Meeting Room}	& \multirow{2}{*}{0}& 2.4					\\
			\cmidrule{4-4}
			204			& ~								& ~					& 5						\\
			\cmidrule{2-4}
			205			& \multirow{2}{*}{Empty Room}	& \multirow{2}{*}{0}& 2.4					\\
			\cmidrule{4-4}
			206			& ~								& ~					& 5						\\
			\bottomrule
		\end{tabular}
		\label{table_group_200_206}
	\end{table*}

	\section{Data Accessibility, Maintenance, and Usage}
	\label{appendix_b}
	\begin{enumerate}
		\item[\textbullet]{%
			\textit{The dataset should be used for academic research purposes only.}
			Readers should not use WiMANS with malicious intent and should not lead to potential negative social impacts with improved WiFi-based multi-user activity sensing.
		}
		\item[\textbullet]{%
			We will create a public repository as the project page of WiMANS to ensure its accessibility and maintenance.
		}
		\item[\textbullet]{%
			Since the dataset is large, we will upload the dataset to cloud storage and provide a link on the project page for readers to download the data.
			%
		}
		\item[\textbullet]{%
			Raw CSI data are saved in ``*.mat'' files, which can be read using SciPy.\\
			The preprocessed data of CSI amplitude are saved in ``*.npy'' files, which can be read using NumPy.\\
			Video data are saved in ``*.mp4'' files, which can be read using PyTorch.\\
			Annotations are saved in the ``annotation.csv'' file, which can be read using Pandas.
		}
		\item[\textbullet]{%
			Details of how to use the dataset and source code are provided in the README.md file.
		}
		\item[\textbullet]{%
			We will provide and maintain all data, source code, and documents on the public project page.
		}
	\end{enumerate}

	\section{Experiments}
	\label{appendix_c}
	Herein, we provide the details of benchmark experiments in WiMANS. 
	\subsection{Implementation Details}
	\label{subsection_implementation_details}
	We utilize WiMANS to benchmark the multi-user sensing performance of 8 WiFi-based and 6 video-based baselines, with respect to human identification, localization, and activity recognition (HAR).
	For each baseline, we implement \textit{three} different models to recognize identities, locations, and activities, separately.
	\begin{enumerate}
		\item{
			When recognizing identities, we identify which users have presented in the environment, regardless of users’ locations and activities.
		}
		\item{
			When recognizing locations, we recognize the location of each user (identity), regardless of user’s activity.
		}
		\item{
			When recognizing activities, we recognize the activity of each user (identity), regardless of user’s location.
		}
	\end{enumerate}
	Overall, the recognition of identities is irrelevant to users’ locations and activities, while the recognition of locations and activities is relevant to user’s identities.

	To perform multi-label classification for identification, localization, and activity recognition, the details of model outputs are described as below:
	\begin{enumerate}
		\item{
			For the recognition of identities, given that it is irrelevant to locations and activities, we encode identity labels in a vector whose length is 6.
			Each element in the vector corresponds to an identity (user). If no user is in the environment, the ground truth is [0, 0, 0, 0, 0, 0].
			If User 1 is in the environment, the ground truth is [1, 0, 0, 0, 0, 0]. If User 2 and 3 are in the environment, the ground truth is [0, 1, 1, 0, 0, 0]. 
			If User 2 to 6 are in the environment, the ground truth is [0, 1, 1, 1, 1, 1].
			The models thereby predict a probability vector. For example, [0.1, 0.5, 0.5, 0.9, 0.0, 0.2] means there is a probability of 0.9 that User 4 is in the environment.
		}
		\item{
			For the recognition of locations, it is relevant to identities, so we first encode each location in a vector whose length is 5, given 5 locations.
			Each element in the vector corresponds to a location.
			If User 1 is at Location A, the ground-truth encoding will be [1, 0, 0, 0, 0].
			If User 2 is not in the environment, the ground-truth encoding will be [0, 0, 0, 0, 0].
			Finally, for all 6 users, we will result in a ground-truth matrix whose dimension is 6$\times$5, corresponding to different locations of different users.
			The models thereby predict a probability matrix.
			For example, [[0.8, 0.2, 0.1, 0.0, 0.0], [0.0, 0.0, 0.1, 0.0, 0.1], […], […], […], […]] means there is a probability of 0.8 that User 1 is at Location A, a probability of 0.1 that User 2 is at Location C, \textit{etc}.
		}
		\item{
			The recognition of activities is similar to location recognition.
			We encode each activity in a vector whose length is 9, given 9 activities.
			If User 1 is performing the first activity, the ground-truth encoding will be [1, 0, 0, 0, 0, 0, 0, 0, 0].
			If User 2 is not in the environment, the ground-truth encoding will be [0, 0, 0, 0, 0, 0, 0, 0, 0].
			Finally, for all 6 users, we will result in a ground-truth matrix whose dimension is 6$\times$9.
			The models thereby predict a probability matrix.
			For example, [[0.4, 0.2, 0.3, 0.0, 0.5, 0.0, 0.0, 0.0, 0.0], […], […], […], […],[…]] means there is a probability of 0.4 that User 1 is doing the first activity, \textit{etc}.
		}
	\end{enumerate}
	Note that we use a sigmoid function (not softmax) following the last layers of models, so the probabilities will not sum up to 1.
	After getting the predicted matrix from models, we use a threshold of 0.5 to determine: (1) whether users are in the environment (to recognize identities), (2) where users are (to recognize locations), (3) and what users are doing (to recognize activities).

	We regard the proportion of correct predictions over all predictions as accuracy.
	For the recognition of locations and activities, we use ``NaN'' to indicate users not presenting in a sample.
	We use top-1 accuracy as metric following previous works \cite{model_ablstm,model_that}, while other metrics (\textit{e.g.}, precision, recall) can be further adopted for evaluation in the future.
	
	%
	%
	%
	Both WiFi-based and video-based models are optimized using Adam \cite{experiment_adam} on a single Nvidia RTX A5000 GPU.
	For fairness, we train all WiFi-based models with the same hyperparameters.
	Similarly, all video-based models are trained with the same hyperparameters.
	Supplement Table \ref{table_train_detail} describes the hyperparameters of all models, regarding the number of epochs, batch size, and learning rate.
	\begin{table*}[t]
		\centering
		\tiny
		\caption{Hyperparameters of WiFi-based and video-base models.}
		\begin{tabular}{@{\quad}l@{\quad}l@{\quad}c@{\quad}c@{\quad}c@{\quad}}
			\toprule
			\tb{Data}						&\tb{Model}						& \tb{Epochs}	& \tb{Batch Size}	& \tb{Learning Rate}		\\
			\midrule
			\multirow{7}{*}{WiFi CSI}		& MLP	\cite{dataset_ntufi}	& 200			& 128				& 0.001								\\
			~								& LSTM \cite{dataset_yousefi}	& 200			& 128				& 0.001									\\
			~								& CNN-1D \cite{model_cnn_1d}	& 200			& 128				& 0.001									\\
			~								& CNN-2D \cite{model_cnn_2d}	& 200			& 128				& 0.001								\\
			~								& CLSTM \cite{model_clstm}		& 200			& 128				& 0.001									\\
			~								& ABLSTM \cite{model_ablstm}	& 200			& 128				& 0.001									\\
			~								& THAT \cite{model_that}		& 200			& 128				& 0.001									\\
			\midrule
			\multirow{6}{*}{Video}			& ResNet	\cite{model_resnet}	& 20		& 8				& 0.0001									\\
			~								& S3D \cite{model_s3d}			& 20		& 8				& 0.0001									\\
			~								& MViT-v1 \cite{model_mvit_v1}	& 20		& 8				& 0.0001									\\
			~								& MViT-v2 \cite{model_mvit_v2}	& 20		& 8				& 0.0001									\\
			~								& Swin-T \cite{model_swin}		& 20		& 8				& 0.0001									\\
			~								& Swin-S \cite{model_swin}		& 20		& 8				& 0.0001									\\
			\bottomrule
		\end{tabular}
		\label{table_train_detail}
	\end{table*}

	\begin{table*}[h!]
		\centering
		\tiny
		\caption{Training time (s) of WiFi-based models.}
		\begin{tabular}{l *{9}{c} }
			\toprule
			\multirow{2}*{\tb{Model}}		& \multicolumn{3}{c}{\tb{Classroom}}	& \multicolumn{3}{c}{\bf{Meeting Room}}		& \multicolumn{3}{c}{\tb{Empty Room}}	\vspace{-0.8mm} \\
			\cmidrule(lr){2-4}\cmidrule(lr){5-7}\cmidrule(lr){8-10}
			~								& Identity	& Location	& Activity 		& Identity	& Location	& Activity			& Identity	& Location	& Activity		\\ 
			\midrule
			\rowcolor[HTML]{EFEFEF} \multicolumn{10}{c}{2.4 GHz}\hspace{0mm}\vspace{1mm}\\
			ST-RF \cite{dataset_yousefi}	& 0.50 		& 1.38 		& 2.67 			& 0.45 		& 1.43 		& 2.65 				& 0.48 		& 1.32 		& 2.67			\\
			MLP	\cite{dataset_ntufi}		& 328.21 	& 218.18 	& 218.29 		& 324.87 	& 215.54 	& 215.76 			& 325.78 	& 216.93 	& 217.99		\\
			LSTM \cite{dataset_yousefi}		& 220.69 	& 220.78 	& 222.96 		& 210.46 	& 215.57 	& 203.69 			& 203.68 	& 216.18 	& 201.99		\\
			CNN-1D \cite{model_cnn_1d}		& 251.78 	& 253.39	& 236.41 		& 234.90 	& 232.07 	& 242.35			& 232.36 	& 241.93	 & 242.32		\\
			CNN-2D \cite{model_cnn_2d}		& 195.85 	& 396.38 	& 389.99 		& 195.38 	& 391.89 	& 388.63 			& 195.21 	& 391.90 	& 388.73		\\
			CLSTM \cite{model_clstm}		& 293.27 	& 291.63 	& 298.63 		& 290.02 	& 289.35 	& 296.96 			& 289.59 	& 289.90 	& 299.49		\\
			ABLSTM \cite{model_ablstm}		& 346.27 	& 264.54 	& 278.64 		& 348.04 	& 262.73 	& 265.65 			& 363.64 	& 262.34 	& 276.06		\\
			THAT \cite{model_that}			& 503.49 	& 703.98	& 669.62 		& 503.55 	& 705.38 	& 669.35 			& 503.63 	& 708.97 	& 705.10		\\
			\midrule
			\rowcolor[HTML]{EFEFEF} \multicolumn{10}{c}{5 GHz}\hspace{0mm}\vspace{1mm}\\
			ST-RF \cite{dataset_yousefi}	& 0.45		& 1.41 		& 2.64 			& 0.39 		& 1.35 		& 2.62 				& 0.45 		& 1.35 		& 2.62 			\\
			MLP	\cite{dataset_ntufi}		& 324.85 	& 218.32 	& 218.36 		& 324.80 	& 215.50 	& 219.36 			& 325.54 	& 216.68 	& 218.54 		\\
			LSTM \cite{dataset_yousefi}		& 223.66 	& 216.66 	& 218.19 		& 210.65 	& 215.65 	& 210.63 			& 203.61 	& 214.21 	& 201.78 		\\
			CNN-1D \cite{model_cnn_1d}		& 246.66 	& 247.69 	& 234.30 		& 232.01 	& 232.18 	& 242.21 			& 232.28 	& 241.77 	& 242.24 		\\
			CNN-2D \cite{model_cnn_2d}		& 196.21 	& 394.83	& 390.30 		& 195.53 	& 396.43 	& 388.83 			& 195.40 	& 391.98 	& 388.92 		\\
			CLSTM \cite{model_clstm}		& 292.95 	& 291.24 	& 298.70 		& 289.24 	& 289.38 	& 297.44 			& 289.40 	& 289.90 	& 299.41 		\\
			ABLSTM \cite{model_ablstm}		& 347.03 	& 264.35 	& 278.61 		& 348.96 	& 263.95 	& 266.23 			& 363.44 	& 262.41 	& 277.09 		\\
			THAT \cite{model_that}			& 503.54 	& 706.03 	& 668.97 		& 503.33 	& 705.04 	& 668.76 			& 502.49 	& 705.63 	& 705.36 		\\
			\midrule
			\rowcolor[HTML]{EFEFEF} \multicolumn{10}{c}{2.4 / 5 GHz}\hspace{0mm}\vspace{1mm}\\
			ST-RF \cite{dataset_yousefi}	& 1.21 		& 3.35 		& 6.46 			& 1.22 		& 3.47 		& 6.75 				& 1.23 		& 3.44 		& 6.21 			\\
			MLP	\cite{dataset_ntufi}		& 648.46 	& 426.99 	& 429.24 		& 649.11 	& 426.36 	& 428.86 			& 648.67 	& 425.78 	& 426.51 		\\
			LSTM \cite{dataset_yousefi}		& 418.46 	& 412.03 	& 413.80 		& 400.79 	& 418.20 	& 397.70 			& 399.42 	& 405.88 	& 397.00 		\\
			CNN-1D \cite{model_cnn_1d}		& 451.40 	& 454.69 	& 457.25 		& 454.15 	& 455.12 	& 494.15 			& 451.73 	& 465.62 	& 500.57 		\\
			CNN-2D \cite{model_cnn_2d}		& 388.74 	& 781.96 	& 775.43 		& 388.86 	& 784.90 	& 775.74 			& 388.85 	& 781.25 	& 775.80 		\\
			CLSTM \cite{model_clstm}		& 577.90 	& 573.85 	& 588.25 		& 571.78 	& 573.81 	& 588.77 			& 571.95 	& 573.86 	& 589.20 		\\
			ABLSTM \cite{model_ablstm}		& 688.71 	& 521.15 	& 521.30 		& 699.94 	& 519.12 	& 527.07 			& 716.97 	& 520.14 	& 535.97 		\\
			THAT \cite{model_that}			& 984.13 	& 1374.70 	& 1311.47 		& 983.85 	& 1379.78 	& 1370.29 			& 983.33 	& 1385.42 	& 1375.65		\\
			\bottomrule
		\end{tabular}
		\label{table_result_csi_train_time}
	\end{table*}

	\subsection{Training Time}
	\label{subsection_training_time}
	To compare the training efficiency of WiFi-based models and video-based models, Supplement Table \ref{table_result_csi_train_time} shows the training time of WiFi-based models, and Supplement Table \ref{table_result_video_train_time} shows the training time of video-based models.
	WiFi-based models are more efficient in terms of training time.

	\begin{table*}[htbp]
		\centering
		\tiny
		\caption{Training time (s) of video-based models.}
		\begin{tabular}{l *{9}{c} }
			\toprule
			\multirow{2}*{\tb{Model}}		& \multicolumn{3}{c}{\tb{Classroom}}	& \multicolumn{3}{c}{\bf{Meeting Room}}		& \multicolumn{3}{c}{\tb{Empty Room}}	\vspace{-0.8mm} \\
			\cmidrule(lr){2-4}\cmidrule(lr){5-7}\cmidrule(lr){8-10}
			~								& Identity	& Location	& Activity 		& Identity	& Location	& Activity			& Identity	& Location	& Activity		\\ 
			\midrule
			ResNet	\cite{model_resnet}		& 3092.13	& 3073.14	& 3064.60		& 3088.87	& 3076.85	& 3069.49			& 3086.77	& 3072.31	& 3069.00		\\
			S3D \cite{model_s3d}			& 3045.62	& 2907.95	& 3009.25		& 2860.61 	& 2949.98	& 2967.23			& 2855.56 	& 2899.53 	& 2894.53		\\
			MViT-v1 \cite{model_mvit_v1}	& 2941.92	& 2928.77	& 2929.15		& 2932.78	& 2927.66	& 2929.29			& 3043.19	& 3046.30	& 3084.37		\\
			MViT-v2 \cite{model_mvit_v2}	& 3929.51	& 3924.34	& 3923.82		& 3915.39	& 3908.96	& 3903.06			& 4027.44	& 3985.62	& 3951.18		\\
			Swin-T \cite{model_swin}		& 2882.28	& 2865.90	& 2866.36		& 2865.69	& 2862.27	& 2862.22			& 2682.28	& 2862.77	& 2861.99		\\
			Swin-S \cite{model_swin}		& 4767.45	& 4757.67	& 4759.70		& 4757.36	& 4754.65	& 4754.16			& 4752.92	& 4755.71	& 4352.64		\\
			\bottomrule
		\end{tabular}%
		\label{table_result_video_train_time}
	\end{table*}%

	\subsection{Testing Time}
	\label{subsection_testing_time}
	To compare the testing efficiency of WiFi-based models and video-based models, Supplement Table \ref{table_result_csi_test_time} shows the testing time of WiFi-based models, and Supplement Table \ref{table_result_video_test_time} shows the testing time of video-based models.
	WiFi-based models are more efficient in terms of testing time.

	\begin{table*}[htbp]
		\centering
		\tiny
		\caption{Testing time (s) of WiFi-based models.}
		\begin{tabular}{l *{9}{c} }
			\toprule
			\multirow{2}*{\tb{Model}}		& \multicolumn{3}{c}{\tb{Classroom}}	& \multicolumn{3}{c}{\bf{Meeting Room}}		& \multicolumn{3}{c}{\tb{Empty Room}}	\vspace{-0.8mm} \\
			\cmidrule(lr){2-4}\cmidrule(lr){5-7}\cmidrule(lr){8-10}
			~								& Identity	& Location	& Activity 		& Identity	& Location	& Activity			& Identity	& Location	& Activity		\\ 
			\midrule
			\rowcolor[HTML]{EFEFEF} \multicolumn{10}{c}{2.4 GHz}\hspace{0mm}\vspace{1mm}\\
			ST-RF \cite{dataset_yousefi}	& 0.0022 	& 0.0069 	& 0.0108 		& 0.0021 	& 0.0067 	& 0.0111 			& 0.0023 	& 0.0062 	& 0.0113 		\\
			MLP	\cite{dataset_ntufi}		& 0.1271 	& 0.1119 	& 0.1113 		& 0.1268 	& 0.1110 	& 0.1112 			& 0.1270 	& 0.1109 	& 0.1116 		\\
			LSTM \cite{dataset_yousefi}		& 0.1233 	& 0.1227 	& 0.1224 		& 0.1231 	& 0.1233 	& 0.1225 			& 0.1232 	& 0.1229 	& 0.1224 		\\
			CNN-1D \cite{model_cnn_1d}		& 0.1379 	& 0.1390 	& 0.1392 		& 0.1390 	& 0.1395 	& 0.1395 			& 0.1383 	& 0.1393 	& 0.1406 		\\
			CNN-2D \cite{model_cnn_2d}		& 0.1739 	& 0.1703 	& 0.1744 		& 0.1756 	& 0.1702 	& 0.1738 			& 0.1738 	& 0.1702 	& 0.1738 		\\
			CLSTM \cite{model_clstm}		& 0.1384 	& 0.1384 	& 0.1362 		& 0.1382 	& 0.1384 	& 0.1358 			& 0.1387 	& 0.1384 	& 0.1358 		\\
			ABLSTM \cite{model_ablstm}		& 0.1508 	& 0.1438 	& 0.1383 		& 0.1508 	& 0.1433 	& 0.1386 			& 0.1500 	& 0.1443 	& 0.1386 		\\
			THAT \cite{model_that}			& 0.1943 	& 0.1906 	& 0.1944 		& 0.1942 	& 0.1908 	& 0.1939 			& 0.1950 	& 0.1921 	& 0.1905		\\
			\midrule
			\rowcolor[HTML]{EFEFEF} \multicolumn{10}{c}{5 GHz}\hspace{0mm}\vspace{1mm}\\
			ST-RF \cite{dataset_yousefi}	& 0.0021 	& 0.0068 	& 0.0110 		& 0.0021 	& 0.0066 	& 0.0111 			& 0.0022 	& 0.0063 	& 0.0111 		\\
			MLP	\cite{dataset_ntufi}		& 0.1263 	& 0.1109 	& 0.1112 		& 0.1271 	& 0.1111 	& 0.1116 			& 0.1269 	& 0.1108 	& 0.1128 		\\
			LSTM \cite{dataset_yousefi}		& 0.1227 	& 0.1231 	& 0.1226 		& 0.1231 	& 0.1235 	& 0.1226 			& 0.1232 	& 0.1226 	& 0.1223 		\\
			CNN-1D \cite{model_cnn_1d}		& 0.1396 	& 0.1394 	& 0.1411 		& 0.1391 	& 0.1381 	& 0.1395 			& 0.1400 	& 0.1398 	& 0.1402 		\\
			CNN-2D \cite{model_cnn_2d}		& 0.1737 	& 0.1701 	& 0.1752 		& 0.1751 	& 0.1697 	& 0.1736 			& 0.1738 	& 0.1704 	& 0.1737 		\\
			CLSTM \cite{model_clstm}		& 0.1384 	& 0.1380 	& 0.1366 		& 0.1383 	& 0.1381 	& 0.1354 			& 0.1385 	& 0.1383 	& 0.1356 		\\
			ABLSTM \cite{model_ablstm}		& 0.1503 	& 0.1438 	& 0.1385 		& 0.1506 	& 0.1431 	& 0.1388 			& 0.1503 	& 0.1440 	& 0.1388 		\\
			THAT \cite{model_that}			& 0.1940 	& 0.1906 	& 0.1941 		& 0.1938 	& 0.1911 	& 0.1953 			& 0.1954 	& 0.1914 	& 0.1907		\\
			\midrule
			\rowcolor[HTML]{EFEFEF} \multicolumn{10}{c}{2.4 / 5 GHz}\hspace{0mm}\vspace{1mm}\\
			ST-RF \cite{dataset_yousefi}	& 0.0041 	& 0.0117 	& 0.0194 		& 0.0039 	& 0.0112 	& 0.0222 			& 0.0042 	& 0.0108 	& 0.0185 		\\
			MLP	\cite{dataset_ntufi}		& 0.2664 	& 0.2186 	& 0.2198 		& 0.2655 	& 0.2202 	& 0.2199 			& 0.2667 	& 0.2205 	& 0.2209 		\\
			LSTM \cite{dataset_yousefi}		& 0.2479 	& 0.2485 	& 0.2482 		& 0.2495 	& 0.2507 	& 0.2484 			& 0.2474 	& 0.2486 	& 0.2477 		\\
			CNN-1D \cite{model_cnn_1d}		& 0.2848 	& 0.2860 	& 0.2859 		& 0.2834 	& 0.2855 	& 0.2864			& 0.2855 	& 0.2855 	& 0.2837 		\\
			CNN-2D \cite{model_cnn_2d}		& 0.3635 	& 0.3534 	& 0.3589 		& 0.3594 	& 0.3544 	& 0.3601 			& 0.3617 	& 0.3530 	& 0.3603 		\\
			CLSTM \cite{model_clstm}		& 0.2817 	& 0.2809 	& 0.2773 		& 0.2809 	& 0.2810 	& 0.2745 			& 0.2805 	& 0.2806 	& 0.2759 		\\
			ABLSTM \cite{model_ablstm}		& 0.3032 	& 0.2881 	& 0.2792		& 0.3041 	& 0.2876 	& 0.2790 			& 0.3034 	& 0.2895 	& 0.2788		\\
			THAT \cite{model_that}			& 0.3866 	& 0.3791 	& 0.3874 		& 0.3868 	& 0.3809 	& 0.3791 			& 0.3884 	& 0.3813 	& 0.3794		\\
			\bottomrule
		\end{tabular}%
		\label{table_result_csi_test_time}
	\end{table*}%

	\begin{table*}[!h]
		\centering
		\tiny
		\caption{Testing time (s) of video-based models.}
		\begin{tabular}{l *{9}{c} }
			\toprule
			\multirow{2}*{\tb{Model}}		& \multicolumn{3}{c}{\tb{Classroom}}	& \multicolumn{3}{c}{\bf{Meeting Room}}		& \multicolumn{3}{c}{\tb{Empty Room}}	\vspace{-0.8mm} \\
			\cmidrule(lr){2-4}\cmidrule(lr){5-7}\cmidrule(lr){8-10}
			~								& Identity	& Location	& Activity 		& Identity	& Location	& Activity			& Identity	& Location	& Activity		\\ 
			\midrule
			ResNet	\cite{model_resnet}		& 14.8304 	& 14.7269 	& 14.7159 		& 14.8047 	& 14.7558 	& 14.7458 			& 14.8281 	& 14.7277 	& 14.7442		\\
			S3D \cite{model_s3d}			& 6.6145 	& 5.8255 	& 6.0780 		& 6.5825 	& 6.6304 	& 6.6416 			& 6.6438 	& 6.3073 	& 6.5706		\\
			MViT-v1 \cite{model_mvit_v1}	& 15.8329 	& 15.7281 	& 15.7698 		& 15.6817 	& 15.7054 	& 15.7075 			& 15.6164 	& 15.5965 	& 15.6433		\\
			MViT-v2 \cite{model_mvit_v2}	& 19.7866 	& 19.9091 	& 19.8755 		& 19.6764 	& 19.7066 	& 19.6765 			& 19.6178 	& 19.5722 	& 19.6451		\\
			Swin-T \cite{model_swin}		& 16.7442 	& 16.8500 	& 16.8160 		& 16.7748 	& 16.7784 	& 16.7767 			& 16.7945 	& 16.7911 	& 16.7976		\\
			Swin-S \cite{model_swin}		& 27.5964 	& 27.4079 	& 27.4720 		& 27.4368 	& 27.4196 	& 27.4380 			& 27.4264 	& 27.5135 	& 27.5100 		\\
			\bottomrule
		\end{tabular}%
		\label{table_result_video_test_time}
	\end{table*}%

	\section{Further Discussions}
	\label{appendix_d}
	\subsection{Ethical Consideration}
	\begin{enumerate}
		\item[\textbullet]{
			Human subject study in WiMANS has been reviewed and approved by institutional ethics committee (IRB).
			Prior to data collection, each subject has been given the details of WiMANS and signed the consent form regarding safety, privacy, and releasing identifiable information.
			The activities in WiMANS are common in daily life (\textit{e.g.}, Walking) to minimize participant risks.
			We create aliases for all subjects (\textit{e.g.}, User 1) and blur their faces in videos to anonymize personally identifiable information.
			
		}
		\item[\textbullet]{
			Improved human identification, localization, and activity recognition may lead to known risks.
			Readers should not misuse WiMANS and should not lead to potential negative social impact.
			
		}
		\item[\textbullet]{
			WiMANS should be used for academic research purposes only.
			It is forbidden to use WiMANS for commercial surveillance systems or other systems that can potentially harm societies or individuals.
			
		}
	\end{enumerate}

	\subsection{Questions and Answers}
	\begin{enumerate}
		\item[\textbullet]{
			\textbf{What are the practical implications and potential applications of multi-user activity sensing in real-world scenarios?}

			WiFi-based multi-user activity sensing can serve as a fundamental component in real-world scenarios as diverse as security monitoring, smart home, and healthcare.
			In security monitoring, traditional single-user approaches may only detect dangers caused by each individual person, while multi-user sensing can recognize dangers in crowd.
			In smart homes, using single-user approach may not satisfy different users simultaneously, while multi-user approaches can provide customized services for different users at the same time.
			Similarly, healthcare based on multi-user sensing can monitor the physical conditions of different users simultaneously.
		}
		\item[\textbullet]{
			\textbf{What are the potential challenges of WiFi-based multi-user activity sensing?}

			The first challenge in multi-user sensing will be the increasing occlusion.
			Because the principle of WiFi-based human sensing is that human motions lead to interference with wireless signals due to reflection and/or refraction, multiple users in an environment may be occluded by each other, challenging the resolution of human sensing approaches.

			The second challenge in multi-user sensing will be the mutual interference between users.
			Compared with single-user sensing which suffers from excessive environmental noise, multi-user sensing needs to further disentangle features of different users to perform effective human sensing.
			
		}
		\item[\textbullet]{
			\textbf{Why are single-user models not applicable to multi-user scenarios?}

			WiFi-based multi-user sensing is more practical yet challenging because CSI is impacted by not only environmental noise but also the mutual interference between users.
			This requires further studies on learning distinguishable human features for sensing.
			
			Existing single-user models formulate the recognition of identities, locations, and activities as single-label classifications, where models only need to predict an output with the highest probability.
			Compared to single-user sensing, multi-user sensing requires multi-label classifications, which need to predict multiple outputs.
			In such multi-label classifications, how to disentangle representative features (of different users) for different outputs is more challenging.

			Empirically, we exploit existing single-user models for WiFi-based multi-user sensing, as shown in Table 2.
			The results show that there is still vast room for improvement.
			Therefore, multi-user models will worth further exploration.
		}
	\end{enumerate}

	\section{Datasheet: Dataset Documentation and Intended Uses}
	\label{appendix_e}
	The following questions are copied from ``Datasheets for Datasets'' \cite{appendix_datasheet}.

	\subsection{Motivation}
	\begin{enumerate}
		\item[\textbullet]{
			\textbf{For what purpose was the dataset created?} Was there a specific task in mind? Was there a specific gap that needed to be filled? Please provide a description.

			WiFi-based human sensing has emerged as a promising solution to analyze user behaviors in a non-intrusive and device-free manner.
			However, most existing works only perform single-user sensing, and all public datasets collect single-user samples only.
			Without corresponding datasets, it is difficult to perform WiFi-based multi-user activity sensing.
			Several recent studies have begun to investigate WiFi-based multi-user sensing, but their datasets are limited in scope and not published.
			This dataset, WiMANS, is created to bridge this gap and facilitate new research on multi-user activity sensing based on WiFi.
			To the best of our knowledge, this is the first dataset which includes multiple users in each sample based on WiFi Channel State Information (CSI).
		}
		\item[\textbullet]{%
			\textbf{Who created the dataset (\textit{e.g.}, which team, research group) and on behalf of which entity (\textit{e.g.}, company, institution, organization)?}

			The authors. To conform with the double blind review policy, we will update the answer to this question after the review process.
		}
		\item[\textbullet]{%
			\textbf{Who funded the creation of the dataset?} If there is an associated grant, please provide the name of the grantor and the grant name and number.

			No direct funding and no conflict of interest. 
		}
	\end{enumerate}

	\subsection{Composition}
	\begin{enumerate}
		\item[\textbullet]{%
			\textbf{What do the instances that comprise the dataset represent (\textit{e.g.}, documents, photos, people, countries)?} Are there multiple types of instances (\textit{e.g.}, movies, users, and ratings; people and interactions between them; nodes and edges)? Please provide a description.

			There are two types of instances: WiFi CSI instances and video instances.
			WiFi CSI records the variations of wireless signals, while videos are taken with a monitor camera to record the activities.
		}
		\item[\textbullet]{%
			\textbf{How many instances are there in total (of each type, if appropriate)?}

			There are 11286 instances in total, consisting of both WiFi CSI and synchronized videos.
			Hence, there are 11286 CSI instances and 11286 video instances.
			WiMANS incorporates varying numbers of users in all the instances.
			There are 594 instances for 0 users, 3564 instances for 1 user, 1782 instances for 2 users, 1782 instances for 3 users, 1782 instances for 4 users, and 1782 instances for 5 users.
			From another aspect, there are 3 environments in WiMANS, and 3762 instances for each environment.
		}
		\item[\textbullet]{%
			\textbf{Does the dataset contain all possible instances or is it a sample (not necessarily random) of instances from a larger set?} If the dataset is a sample, then what is the larger set? Is the sample representative of the larger set (\textit{e.g.}, geographic coverage)? If so, please describe how this representativeness was validated/verified. If it is not representative of the larger set, please describe why not (\textit{e.g.}, to cover a more diverse range of instances, because instances were withheld or unavailable).

			The entire dataset is presented (it is not a sample of another dataset).
			%
			%
			We select 9 daily activities that exist in the real world to build up the dataset, and these activities are common and representative in daily life, following the previous works \cite{dataset_yousefi,dataset_widar3,dataset_operanet}.
			Moreover, different numbers of users perform these activities independently yet simultaneously, resulting in various combinations of activities.
			We will extend to more activities in the future. 
		}
		\item[\textbullet]{%
			\textbf{What data does each instance consist of?} ``Raw'' data (\textit{e.g.}, unprocessed text or images) or features? In either case, please provide a description.

			We include raw WiFi CSI instances in WiMANS.
			Each CSI instance is a matrix of complex values with the dimension of 3000$\times$3$\times$30$\times$30.
			We also preprocess raw WiFi CSI instances by calculating their amplitude and provide the amplitude matrices in the dataset.
			For videos, WiMANS contains synchronized videos, and the dimension of each video is 90$\times$3$\times$1920$\times$1080.
		}
		\item[\textbullet]{%
			\textbf{Is there a label or target associated with each instance?} If so, please provide a description.
			
			%
			%
			We label each sample as ``act\_$<$group$>$\_$<$sample$>$'', where ``$<$group$>$'' is the user group index and ``$<$sample$>$'' is the sample index.
			We associate each label with the environment, WiFi band, number of users, user identities, locations, and activities, to enable different tasks.
			The environments include: (1) Classroom, (2) Meeting Room, (3) Empty Room.
			The numbers of users range from 0 to 5.
			The user identities are: User 1, User 2, User 3, User 4, User 5, and User 6.
			The locations are: A, B, C, D, and E.
			The activities are: (1) Nothing, (2) Walking, (3) Rotation, (4) Jumping, (5) Waving, (6) Lying Down, (7) Picking Up, (8) Sitting Down, (9) Standing Up.
			Note that the videos share the same labels with CSI for reference and unexplored tasks.
		}
		\item[\textbullet]{%
			\textbf{Is any information missing from individual instances?} If so, please provide a description, explaining why this information is missing (\textit{e.g.}, because it was unavailable). This does not include intentionally removed information, but might include, \textit{e.g.}, redacted text.

			No.
		}
		\item[\textbullet]{%
			\textbf{Are relationships between individual instances made explicit (\textit{e.g.}, users' movie ratings, social network links)?} If so, please describe how these relationships are made explicit.

			No.
		}
		\item[\textbullet]{%
			\textbf{Are there recommended data splits (\textit{e.g.}, training, development/ validation, testing)?} If so, please provide a description of these splits, explaining the rationale behind them.

			We randomly split our dataset into a training set (80\%) and a test set (20\%), following the previous works \cite{model_ablstm,model_that}.
		}
		\item[\textbullet]{%
			\textbf{Are there any errors, sources of noise, or redundancies in the dataset?} If so, please provide a description.

			Ideally, each CSI sample consists of 3000 time steps.
			However, there are missing time steps in WiFi CSI owing to packet loss, and the average packet loss rates are 4.52\% and 2.31\% for 2.4 GHz band and 5 GHz band.
			On average, 2.4 GHz instances have 2864.39 time steps, while 5 GHz instances have 2930.72 time steps.
			Given that such numbers of missing steps are inevitable yet acceptable, we can solve this issue by zero padding.

			Regarding the noise in WiMANS, CSI typically intertwines with environmental noise, because WiFi is initially designed for wireless communication rather than sensing.
			Meanwhile, WiFi-based models aim to extract human sensing features from CSI by getting rid of noise.
			Therefore, such noise in the dataset is reasonable.

			As mentioned before, we include both raw CSI and preprocessed CSI amplitude in WiMANS.
			CSI amplitude instances can be calculated based on raw CSI instances, which is a sort of redundancy.
		}
		\item[\textbullet]{%
			\textbf{Is the dataset self-contained, or does it link to or otherwise rely on external resources (\textit{e.g.}, websites, tweets, other datasets)?} If it links to or relies on external resources, a) are there guarantees that they will exist, and remain constant, over time; b) are there official archival versions of the complete dataset (\textit{i.e.}, including the external resources as they existed at the time the dataset was created); c) are there any restrictions (\textit{e.g.}, licenses, fees) associated with any of the external resources that might apply to a dataset consumer? Please provide descriptions of all external resources and any restrictions associated with them, as well as links or other access points, as appropriate.

			Yes, the dataset is self-contained.
		}
		\item[\textbullet]{%
			\textbf{Does the dataset contain data that might be considered confidential (\textit{e.g.}, data that is protected by legal privilege or by doctor–patient confidentiality, data that includes the content of individuals' non-public communications)?} If so, please provide a description.

			No.
		}
		\item[\textbullet]{%
			\textbf{Does the dataset contain data that, if viewed directly, might be offensive, insulting, threatening, or might otherwise cause anxiety?} If so, please describe why.

			No.
		}
		\item[\textbullet]{%
			\textbf{Does the dataset identify any subpopulations (\textit{e.g.}, by age, gender)?} If so, please describe how these subpopulations are identified and provide a description of their respective distributions within the dataset.

			We describe the statistics of participants, including their genders, ages, weights, heights, and body mass indexes in Supplement Table \ref{table_users}.
		}
		\item[\textbullet]{%
			\textbf{Is it possible to identify individuals (\textit{i.e.}, one or more natural persons), either directly or indirectly (i.e., in combination with other data) from the dataset?} If so, please describe how.

			We create aliases for all subjects (e.g., User 1) and blur their faces in videos to anonymize personally identifiable information.
			It is possible to identify user aliases by analyzing WiFi CSI and videos.
			Volunteers have signed the consent form regarding the release of identifiable information.
		}
		\item[\textbullet]{%
			\textbf{Does the dataset contain data that might be considered sensitive in any way (\textit{e.g.}, data that reveals race or ethnic origins, sexual orientations, religious beliefs, political opinions or union memberships, or locations; financial or health data; biometric or genetic data; forms of government identification, such as social security numbers; criminal history)?} If so, please provide a description.

			The dataset provides WiFi CSI and videos, containing the information of (anonymized) identities, locations, and activities.
		}
	\end{enumerate}

	\subsection{Collection Process}
	\begin{enumerate}
		\item[\textbullet]{%
			\textbf{How was the data associated with each instance acquired?} Was the data directly observable (\textit{e.g.}, raw text, movie ratings), reported by subjects (\textit{e.g.}, survey responses), or indirectly inferred/derived from other data (\textit{e.g.}, part-of-speech tags, model-based guesses for age or language)? If the data was reported by subjects or indirectly inferred/derived from other data, was the data validated/verified? If so, please describe how.

			We collect WiFi CSI using a WiFi transmitter and a WiFi receiver.
			These devices are equipped with Intel 5300 Network Interface Cards and have the Linux 802.11n CSI tool \cite{experiment_80211} installed.
			These devices work in the monitor mode on channel 12 for the 2.4 GHz band, and on channel 64 for the 5 GHz band.
			Meanwhile, we collect videos using a monitor camera.
			WiFi CSI cannot be interpreted directly, while videos are observable.
			More details are provided in Section 3.1 of the main paper.	
		}
		\item[\textbullet]{%
			\textbf{What mechanisms or procedures were used to collect the data (\textit{e.g.}, hardware apparatuses or sensors, manual human curation, software programs, software APIs)?} How were these mechanisms or procedures validated?

			We employ the transmitter and receiver to collect a CSI sample in 3 steps: (1) The receiver listens to a WiFi channel and logs the CSI of all packets it receives; (2) The transmitter sends packets to the WiFi channel, and the users simultaneously perform designated activities; (3) The receiver stops logging and listening.
			Simultaneously, the monitor camera records the videos.
			
			The dataset is collected in 11286 sessions, where we collect CSI data sample by sample, while recording long videos at the same time.
			Between the collection of each two samples, we will first stop the collection of previous sample and then announce a new index to users, after which we will start the collection of the next sample. 
			Therefore, we can ensure the purity of each sample, not containing nearby activities.

			After data collection, we use the timestamps in CSI samples to crop long videos.
			Because the frame rate of videos is 30 Hz while the sample rate of CSI is 1000 Hz, we can synchronize WiFi CSI and video samples in 16.67 (=1000/30/2) ms, which is unavoidable and tolerable (for 3-second samples).
		}
		\item[\textbullet]{%
			\textbf{If the dataset is a sample from a larger set, what was the sampling strategy (\textit{e.g.}, deterministic, probabilistic with specific sampling probabilities)?}

			WiMANS is not a sample from a larger set. 
			We select 9 daily activities to build up the dataset, following the previous works \cite{dataset_yousefi,dataset_widar3,dataset_operanet}.
			Meanwhile, different numbers of users perform these activities independently yet simultaneously, resulting in various combinations of activities.
		}
		\item[\textbullet]{%
			\textbf{Who was involved in the data collection process (\textit{e.g.}, students, crowdworkers, contractors) and how were they compensated (\textit{e.g.}, how much were crowdworkers paid)?}

			6 volunteers act as users in the data collection process.
			%
			%
			They volunteered, and their compensation was to be provided with lunch and dinner.
		}
		\item[\textbullet]{%
			\textbf{Over what timeframe was the data collected?} Does this timeframe match the creation timeframe of the data associated with the instances (\textit{e.g.}, recent crawl of old news articles)? If not, please describe the timeframe in which the data associated with the instances was created.

			According to aforementioned data collection procedures, we instruct users to perform activities and collect data simultaneously.
			Hence, the timeframe can match the activities which we would like to monitor.
		}
		\item[\textbullet]{%
			\textbf{Were any ethical review processes conducted (\textit{e.g.}, by an institutional review board)?} If so, please provide a description of these review processes, including the outcomes, as well as a link or other access point to any supporting documentation.

			Yes.
			Human subject study in WiMANS has been reviewed and approved by the institutional ethics committee (IRB).
		}
		\item[\textbullet]{%
			\textbf{Did you collect the data from the individuals in question directly, or obtain it via third parties or other sources (\textit{e.g.}, websites)?}

			Yes. We collect data by directly recording WiFi CSI and videos when volunteers are performing activities.
		}
		\item[\textbullet]{%
			\textbf{Were the individuals in question notified about the data collection?} If so, please describe (or show with screenshots or other information) how notice was provided, and provide a link or other access point to, or otherwise reproduce, the exact language of the notification itself.

			Yes.
			We tell them that they need to perform designated activities at designated locations in designated environments according to the scripts we give them.
			Figure 4 in the main paper and Supplement Tables \ref{table_group_1_18} $\sim$ \ref{table_group_200_206} present these scripts and the arrangement of user groups.
		}
		\item[\textbullet]{%
			\textbf{Did the individuals in question consent to the collection and use of their data?} If so, please describe (or show with screenshots or other information) how consent was requested and provided, and provide a link or other access point to, or otherwise reproduce, the exact language to which the individuals consented.

			Yes.\\
			(1) Volunteers consent to take part in WiMANS: A Benchmark Dataset for WiFi-based Multi-user Activity Sensing. \\
			(2) Volunteers understand that their participation is voluntary, and they are free to withdraw at any time, without giving any reason and without their legal rights nor treatment / healthcare being affected.\\
			(3) Volunteers give consent for information and samples collected from them to be used to support other research or in the development of a new test, medication, medical device or treatment by an academic institution or commercial company in the future.\\
			(4) Volunteers understand that data collected from them are a gift donated to the authors and that they will not personally benefit financially if this research leads to an invention and/or the successful developments of a new test, medication treatment, product or service.
		}
		\item[\textbullet]{%
			\textbf{If consent was obtained, were the consenting individuals provided with a mechanism to revoke their consent in the future or for certain uses?} If so, please provide a description, as well as a link or other access point to the mechanism (if appropriate).

			Participants can revoke their consent by contacting the authors.
		}
		\item[\textbullet]{%
			\textbf{Has an analysis of the potential impact of the dataset and its use on data subjects (\textit{e.g.}, a data protection impact analysis) been conducted?} If so, please provide a description of this analysis, including the outcomes, as well as a link or other access point to any supporting documentation.

			No data impact analysis has been carried out explicitly.
			We create aliases for all subjects (e.g., User 1) and blur their faces in videos to anonymize personally identifiable information.
			We do not store the individual names \textit{etc.}, which ensures their identities cannot be correlated.
			As far as we know, the data collection does not endanger any individual because the volunteers only need to perform daily activities.
			Meanwhile, WiMANS is for academic research purposes only.
			Readers should not use WiMANS with malicious intent and should not lead to potential negative social impacts.
		}
	\end{enumerate}

	\subsection{Preprocessing/Cleaning/Labeling}
	\begin{enumerate}
		\item[\textbullet]{%
			\textbf{Was any preprocessing/cleaning/labeling of the data done (\textit{e.g.}, discretization or bucketing, tokenization, part-of-speech tagging, SIFT feature extraction, removal of instances, processing of missing values)?} If so, please provide a description. If not, you may skip the remaining questions in this section.

			We preprocess raw CSI samples by calculating their amplitude.
			We do not process the missing time steps in raw CSI samples, because they can be easily fixed by zero padding.
		}
		\item[\textbullet]{%
			\textbf{Was the ``raw'' data saved in addition to the preprocessed/cleaned/ labeled data (\textit{e.g.}, to support unanticipated future uses)?} If so, please provide a link or other access point to the ``raw'' data.

			Yes. We provide the raw data together with the preprocessed data, using the same link.
		}
		\item[\textbullet]{%
			\textbf{Is the software that was used to preprocess/clean/label the data available?} If so, please provide a link or other access point.

			We use SciPy 1.7.3 and NumPy 1.21.5 to read and preprocess the CSI data.\\
			SciPy: \url{https://scipy.org}\\
			NumPy: \url{https://numpy.org}\\
			We use PyTorch 2.0.1 to read and preprocess the video data.\\
			PyTorch: \url{https://pytorch.org}\\
			We use Pandas 1.4.2 to read the annotation file.\\
			Pandas: \url{https://pandas.pydata.org}\\
			All the above packages are based on Python 3.9.12.\\
			Python: \url{https://www.python.org}
		}
	\end{enumerate}

	\subsection{Uses}
	\begin{enumerate}
		\item[\textbullet]{%
			\textbf{Has the dataset been used for any tasks already?} If so, please provide a description.

			Yes.
			We use WiMANS to conduct benchmark experiments.
			Specifically, we have exploited WiMANS for human identification, human localization, and human activity recognition (HAR).
		}
		\item[\textbullet]{%
			\textbf{Is there a repository that links to any or all papers or systems that use the dataset?} If so, please provide a link or other access point.

			Yes, we will create a repository as the project page to provide all data, source code, and documents.
			To conform with the double blind review policy, we will update the answer to this question after the review process.
		}
		\item[\textbullet]{%
			\textbf{What (other) tasks could the dataset be used for?}

			Given that we also include videos in WiMANS, WiFi-based multi-user pose estimation can be further studied after annotating videos with human body joints.
		}
		\item[\textbullet]{%
			\textbf{Is there anything about the composition of the dataset or the way it was collected and preprocessed/cleaned/labeled that might impact future uses?} For example, is there anything that a dataset consumer might need to know to avoid uses that could result in unfair treatment of individuals or groups (\textit{e.g.}, stereotyping, quality of service issues) or other risks or harms (\textit{e.g.}, legal risks, financial harms)? If so, please provide a description. Is there anything a dataset consumer could do to mitigate these risks or harms?

			WiMANS is for academic research purposes only.
			Readers should not use WiMANS with malicious intent and should not lead to potential negative social impacts.
		}
		\item[\textbullet]{%
			\textbf{Are there tasks for which the dataset should not be used?} If so, please provide a description.

			The dataset should not be used for any illegal tasks.
			The dataset should be used for academic research purposes only.
			%
		}
	\end{enumerate}

	\subsection{Distribution}
	\begin{enumerate}
		\item[\textbullet]{%
			\textbf{Will the dataset be distributed to third parties outside of the entity (\textit{e.g.}, company, institution, organization) on behalf of which the dataset was created?} If so, please provide a description.

			No.
		}
		\item[\textbullet]{%
			\textbf{How will the dataset will be distributed (\textit{e.g.}, tarball on website, API, GitHub)?} Does the dataset have a digital object identifier (DOI)?

			We will make WiMANS publicly available online by creating a repository as the project page.
			Since the dataset is large, we will upload the data to cloud storage and provide a link on the project page for readers to access all the data.
			Source code and documents will be available on the project page as well.
			We will create a DOI for the dataset after the paper is accepted.
		}
		\item[\textbullet]{%
			\textbf{When will the dataset be distributed?}

			After the paper is accepted.
		}
		\item[\textbullet]{%
			\textbf{Will the dataset be distributed under a copyright or other intellectual property (IP) license, and/or under applicable terms of use (ToU)?} If so, please describe this license and/or ToU, and provide a link or other access point to, or otherwise reproduce, any relevant licensing terms or ToU, as well as any fees associated with these restrictions.

			Attribution-NonCommercial-ShareAlike 4.0 License (CC BY-NC-SA 4.0).
		}
		\item[\textbullet]{%
			\textbf{Have any third parties imposed IP-based or other restrictions on the data associated with the instances?} If so, please describe these restrictions, and provide a link or other access point to, or otherwise reproduce, any relevant licensing terms, as well as any fees associated with these restrictions.

			No.
		}
		\item[\textbullet]{%
			\textbf{Do any export controls or other regulatory restrictions apply to the dataset or to individual instances?} If so, please describe these restrictions, and provide a link or other access point to, or otherwise reproduce, any supporting documentation.

			WiMANS is for academic research purposes only.
			Readers should not use WiMANS with malicious intent and should not lead to potential negative social impacts.
		}
	\end{enumerate}

	\subsection{Maintenance}
	\begin{enumerate}
		\item[\textbullet]{
			\textbf{Who will be supporting/hosting/maintaining the dataset?}

			The authors.
			To conform with the double blind review policy, we will update the answer to this question after the review process.
		}
		\item[\textbullet]{%
			\textbf{How can the owner/curator/manager of the dataset be contacted (\textit{e.g.}, email address)?}

			The authors.
			To conform with the double blind review policy, we will update the answer to this question after the review process.
		}
		\item[\textbullet]{%
			\textbf{Is there an erratum?} If so, please provide a link or other access point.

			No.
		}
		\item[\textbullet]{%
			\textbf{Will the dataset be updated (\textit{e.g.}, to correct labeling errors, add new instances, delete instances)?} If so, please describe how often, by whom, and how updates will be communicated to dataset consumers (\textit{e.g.}, mailing list, GitHub)?

			Yes. 
			We plan to continue the development of the dataset.
			Updates will be shared and announced on the repository as soon as they are available.
		}
		
		\item[\textbullet]{%
			\textbf{If the dataset relates to people, are there applicable limits on the retention of the data associated with the instances (\textit{e.g.}, were the individuals in question told that their data would be retained for a fixed period of time and then deleted)?} If so, please describe these limits and explain how they will be enforced.
			
			%
			%
			The volunteers have signed the consent form and understand that data collected from them are a gift donated to the authors.
			They have been made aware that there is no applicable limit on the retention of the data.
		}
		\item[\textbullet]{%
			\textbf{Will older versions of the dataset continue to be supported/hosted/ maintained?} If so, please describe how. If not, please describe how its obsolescence will be communicated to dataset consumers.

			Yes. If we update our dataset, we will maintain the old version and then release the new version (\textit{e.g.}, WiMANS 2.0).
			The project page of WiMANS can maintain the history of versions.
			Any changes regarding the maintenance will be announced in advance on the repository.
		}
		\item[\textbullet]{%
			\textbf{If others want to extend/augment/build on/contribute to the data-set, is there a mechanism for them to do so?} If so, please provide a description. Will these contributions be validated/verified? If so, please describe how. If not, why not? Is there a process for communicating/distributing these contributions to dataset consumers? If so, please provide a description.

			Yes. Contributions to the dataset and benchmarks are highly appreciated.
			Researchers can extend the dataset by annotating videos with human body joints for WiFi-based pose estimation.
			Annotations can be added to the dataset by pulling requests on the repository or though emails to dataset managers.
			All contributions will be properly acknowledged.
		}
	\end{enumerate}

	%
	%
	\bibliographystyle{splncs04}
	\bibliography{main}
\end{document}